%% file: nuNtransitions.tex
\documentclass[reprint,twocolumn,superscriptaddress,amsmath,amssymb,nofootinbib,preprintnumbers]{revtex4-2}

\usepackage{graphicx}
\usepackage{dcolumn}
\usepackage{bm}
\usepackage{upgreek}
\usepackage[mathlines]{lineno}
\usepackage{enumitem}

\newcommand{\dif}{\mathrm{d}}
\newcommand{\LL}{\mathrm{L}}
\newcommand{\RR}{\mathrm{R}}
  
\begin{document}


\title{Search for active--sterile neutrino transitions using Pierre Auger Observatory data}
 


\input{revtex_authorlist.tex}

\date{\today}

\begin{abstract}
We investigate the sensitivity of the Pierre Auger Observatory to physics beyond the Standard Model arising from magnetic-moment–induced transitions between active and heavy sterile neutrinos. Such dipole portal interactions can enhance neutrino–nucleon cross sections above a kinematic threshold set by the sterile neutrino mass, leading to  observable modifications of neutrino detection rates at ultrahigh energies (UHE). We estimate the impact of these interactions on both down-going and Earth-skimming neutrino detection channels, the contrasting responses of which enable discrimination between an enhanced neutrino flux and a modified interaction cross section. Using the non-observation of UHE neutrino candidates, we derive neutrino-flux-dependent constraints with 90\% confidence-level on the transition magnetic moment for sterile neutrino masses in the range 1\,TeV–100\,TeV. Under the assumed flux scenarios, the resulting flavor-independent limits extend existing bounds into previously unexplored parameter space.
\end{abstract}

\maketitle

      
%
\section{Introduction}
\label{intro}

Ultrahigh-energy neutrinos (UHE$\nu$s) are expected from interactions of ultrahigh-energy cosmic rays (UHECRs), either at their acceleration sites or during their propagation through intergalactic space. The resulting  neutrino fluxes remain uncertain, as they depend on several under-constrained astrophysical inputs. Among these, the UHECR composition is  critical, since neutrino production is dominated by the decay of charged pions generated in hadronic interactions. Pion-producing hadrons may be either free protons or nucleons bound within heavier nuclei. Because these nucleons carry approximately the parent nucleus energy divided by its mass number, neutrinos originating from heavy primaries are expected at lower energies than those from proton-dominated compositions. Additional uncertainties stem from the largely unknown properties of the source environments, including photon and gas densities as well as particle confinement times, which govern UHECR energy losses prior to escape. Overall, in mixed-composition scenarios consistent with the spectrum and composition data of the Pierre Auger Observatory, the predicted neutrino fluxes lie well below current detection sensitivities above $1~$EeV~\cite{PierreAuger:2022atd,Aloisio:2013hya,Biehl:2017hnb,AlvesBatista:2018zui,Boncioli:2018lrv,Zhang:2018agl,Condorelli:2022vfa,Muzio:2022bak,Rodrigues:2020pli,Muzio:2023skc,Berat:2024rvf}. Nevertheless, around $E \sim 100~\mathrm{PeV}$, optimistic realizations of these models yield fluxes that are within an order of magnitude of the present IceCube~\cite{BOTNER2005367} and Auger~\cite{PierreAuger:2015eyc} sensitivities, and may already be probed by the recent detection of the KM3-230213A event~\cite{KM3NeT:2025npi,KM3NeT:2025vut}.

The sensitivities of current and planned UHE$\nu$ observatories are evaluated assuming that neutrinos interact with nucleons exclusively through Standard Model (SM) weak interactions, which determine the cross section $\sigma_{\nu\mathrm{n}}(E)$. However, physics beyond the SM (BSM) may lead to observable deviations in $\sigma_{\nu\mathrm{n}}$ at sufficiently high energies. One particularly well-motivated possibility is the existence of a transition magnetic moment, $\mu_{\nu N}$, coupling active neutrinos to heavier sterile states $N$. Although neutrinos carry no electric charge, they can, as any spin-$1/2$ fermion, couple to the electromagnetic field through higher multipole moments generated by quantum loops involving charged particles~\cite{Fujikawa:1980yx,Shrock:1982sc,Giunti:2014ixa}. For a transition between two distinct states $\nu\to N$, the off-diagonal charge form factor is dominated by the magnetic dipole term, proportional to  $\mu_{\nu N}$. Transition moments are moreover the only magnetic moments allowed for Majorana fermions, the diagonal moments of which vanish identically~\cite{Shrock:1982sc}. The resulting effective interaction, including the accompanying $Z$-boson coupling required by gauge invariance, is detailed in Appendix~\ref{sec:app1}. In this scenario, an incident neutrino can upscatter off a nucleon through the dipole interaction, producing an on-shell heavy neutrino via the process $\nu \mathrm{n}\rightarrow N X$, where $X$ denotes the hadronic debris. Once the kinematic threshold $E\gtrsim m_N^2/(2m_{\mathrm n})$ is exceeded, the corresponding cross section rises rapidly with energy because the process is dominated by deep-inelastic scattering on sea quarks, whose densities increase toward small Bjorken-$x$. The heavy neutrino subsequently decays promptly through $N\rightarrow \gamma\nu~\text{or}~N\rightarrow Z\nu$, generating additional electromagnetic and hadronic cascades. Together with the hadronic debris $X$, these decay products contribute to the development of the extensive air shower.

Although many theoretical frameworks predict $\mu_{\nu N}$ values well below current experimental sensitivities for $m_N\gtrsim100~\mathrm{GeV}$, models with family symmetries allow for naturally large transition moments while remaining compatible with phenomenologically acceptable neutrino masses~\cite{Babu:2020ivd}. Hence, searching for magnetic-moment-induced interactions not only tests the SM description of neutrino scattering at extreme energies but also offers a window onto heavy neutral leptons with masses extending into the multi-TeV regime. Yet, rather than focusing on a specific BSM realization, we remain agnostic about the origin of $\mu_{\nu N}$ and treat it in this study as an effective coupling to be constrained experimentally. 

An enhancement in the rate of neutrino–induced events relative to SM expectations may, in a conservative interpretation, originate from a neutrino flux larger than expected. Importantly, the Observatory offers a powerful diagnostic to discriminate between such a flux enhancement and a modification of $\sigma_{\nu\mathrm{n}}$~\cite{Kusenko:2001gj,Anchordoqui:2001cg,Anchordoqui:2010hq}. This discrimination is possible because UHE$\nu$s can be detected either through horizontal air showers initiated in the atmosphere (``down-going'' detection mode~\cite{Capelle:1998zz}) or through up-going showers produced by the decay of $\uptau$ leptons following $\nu_\uptau$ interactions in the crust of the Earth (``Earth-skimming'' detection mode~\cite{Bertou:2001vm}). An increased neutrino flux would raise the event rates in both channels simultaneously. By contrast, an enhancement of the cross section induced by a sizable $\mu_{\nu N}$ would preferentially increase the down-going event rate while suppressing the Earth-skimming channel. Indeed, the atmosphere acts as a thin target, so that the down-going rate grows with $\sigma_{\nu\mathrm{n}}$, while the products of the upscattering initiate air showers similar to SM ones. Earth-skimming $\nu_\uptau$, in contrast, traverse tens to hundreds of kilometers of rock; since $N$ decays back into an active neutrino, each upscattering acts as an additional neutral-current-like interaction that degrades the $\nu_\uptau$ energy without producing the $\uptau$ lepton required for detection.

In this work, we constrain magnetic-moment–induced transitions between active and sterile states with mass between $1\,$TeV and $100\,$TeV that could enhance $\sigma_{\nu\mathrm{n}}$ at extreme energies. The absence of detected events therefore places limits on the size of such effects, translating into forbidden regions in the  $(\mu_{\nu N},\,m_{N})$ plane. To do so, we evaluate the expected number of events for three benchmark  neutrino fluxes spanning the range of current astrophysical uncertainties associated with UHECR composition and source evolution. The resulting exclusion regions in the $(\mu_{\nu N},\,m_{N})$ plane thus provide a model-dependent, yet robust, assessment of the sensitivity of current UHE$\nu$ searches to magnetic-moment-mediated BSM physics.

\section{Active--sterile neutrino transition through the dipole portal}
\label{sec:active-sterile-transition}

In the presence of a magnetic field, massive left-handed (active) neutrinos, $\nu_\LL$, can flip into right-handed (sterile) states, $N_\RR$, through transition magnetic moments, $\mu_{\nu N}$. Such transitions, assumed to be flavor-independent in this study, can occur in the electromagnetic field of a nucleus and therefore provide an additional contribution to neutrino–nucleon scattering beyond SM weak interactions~\cite{Domokos:1996cn}. Gauge invariance of the SM further implies that this process is accompanied by a corresponding contribution mediated by $Z$-boson exchange (see Appendix~\ref{sec:app1}). For a nucleon target with mass $m_\mathrm{n}$, the required kinematic lower bound on the incident neutrino energy $E$ to produce an on-shell heavy sterile neutrino with mass $m_N$ is $E\gtrsim m_N^2/2m_\mathrm{n}\approx 53(m_N/10\,\mathrm{TeV})^2\,\mathrm{PeV}$.  
 
For an UHE$\nu$, the available center-of-mass energy  can reach values sufficient for the upscattering process $\nu \mathrm{n} \to NX$ to take place with $m_N$ up to the deca-TeV scale. The dominant contribution to the total cross section arises from the sea quarks in the deep inelastic scattering (DIS) regime, due to their densities rapidly growing at small Bjorken-$x$. In this regime, the differential cross section for the upscattering process off a nucleon can be expressed as (see Appendix~\ref{sec:app1})
\begin{multline}
    \label{eqn:dsig_dxdy}
    \frac{\dif^{2}\sigma_{\nu \mathrm{n}}}{\dif x\,\dif y}
    = \mu_{\nu N}^2\frac{F(\hat{s},t;m_N)}{2\pi\hat{s}}
      \sum_{i} \mathrm{q}_{i}(x,Q^2
      )\times\strut\\
      \left(\left|\frac{eQ_i}{t}-\frac{gg_i^\mathrm{V}\tan{\theta_\mathrm{W}}}{2\cos\theta_\mathrm{W}(t-m_Z^2)}\right|^2+\frac{(gg_i^\mathrm{A}\tan{\theta_\mathrm{W}})^2}{4\cos^2\theta_\mathrm{W}(t-m_Z^2)^2}\right),
\end{multline}
where $g$ is the electroweak SU(2)$_\LL$ constant, $\theta_\mathrm{W}$ stands for the electroweak mixing angle, $m_Z$ denotes the mass of the $Z$, $eQ_{i}$ ($g_i^\mathrm{V}$) [$g_i^\mathrm{A}$] denotes the electric charge (vector coupling constant) [axial coupling constant] of the $i$th quark, $\mathrm{q}_{i}(x)$ is the corresponding parton distribution function (PDF), and the standard DIS kinematic variables are employed. Specifically,
\begin{eqnarray}
    \label{eqn:DIS}
    \hat{s} &=& 2Exm_\mathrm{n} + x^{2}m_\mathrm{n}^{2}, \\
    t &=& -Q^{2} = -2 Exym_\mathrm{n},
\end{eqnarray}
where $x$ is the Bjorken scaling variable, $y$ is the inelasticity parameter and  $\hat{s}$ represents the center-of-mass energy squared of the neutrino-parton system. The polynomial function $F$ reads as 
\begin{equation}
    \label{eqn:F}
    F(\hat{s},t;m_N)
    = \left(m_N^2(2\hat{s}+t)-2\hat{s}(\hat{s}+t)-m_N^4\right)t,
\end{equation}
where $m_\mathrm{n}$ is neglected in this expression. The PDFs $\mathrm{q}_{i}(x,Q^2)$ are taken from the LHAPDF~6.5.3 library~\cite{Buckley:2014ana}. 

\begin{figure}[t]
\centering
\includegraphics[width=0.5\textwidth]{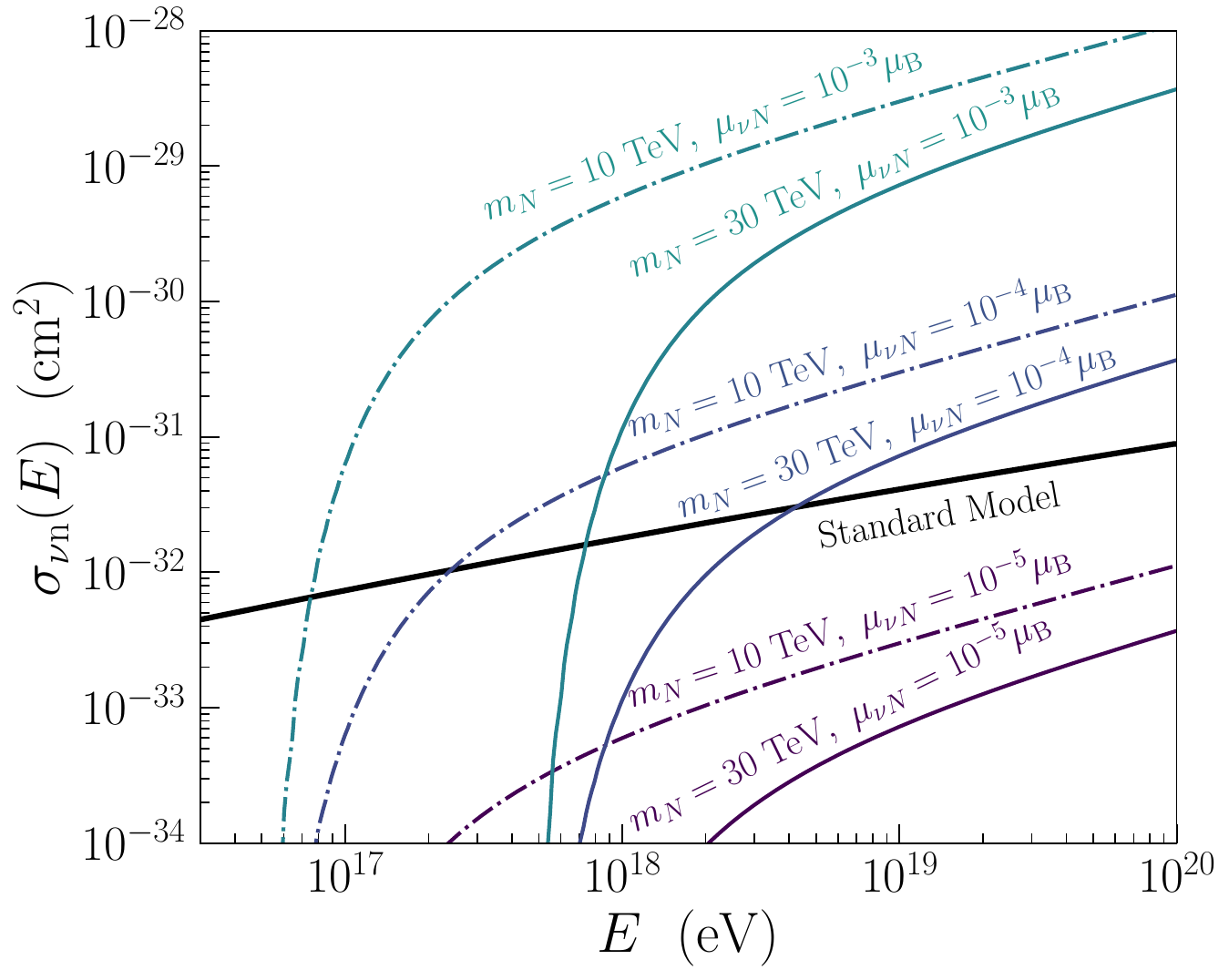}
\caption{Neutrino-nucleon cross section as a function of energy for SM CC interaction (black) and for upscattering into a heavy $N$ state with $m_N=10\,$TeV (dashed lines) and $m_N=30\,$TeV (solid lines). BSM curves are color-coded ranging from dark blue to light blue for $\mu_{\nu N}$ progressing from $10^{-5}\mu_\mathrm{B}$ to $10^{-3}\mu_\mathrm{B}$.}
\label{fig:sigma_vs_E}
\end{figure}

In Fig.~\ref{fig:sigma_vs_E}, we compare the SM charged-current (CC) cross section with that of the $\nu$–$N$ conversion, as a function of the incoming neutrino energy $E$. We consider two representative mass values, $m_N = 10\,\mathrm{TeV}$ (dash-dotted) and $m_N = 30\,\mathrm{TeV}$ (solid). The scaling with  $\mu_{\nu N}^2$, expressed in units of the Bohr magneton $\mu_{\rm B}\simeq 297~\mathrm{GeV}^{-1}$, is apparent. As expected, the cross section rises sharply once the kinematic threshold for heavy sterile neutrino production is crossed. At higher energies it approaches a power-law behavior similar to that of the SM CC interaction (shown as the black curve~\cite{Gandhi:1995tf,Armesto:2007tg,Connolly:2011vc}), since both processes are dominated by DIS on small-$x$ partons. The cross sections shown remain consistent with IceCube measurements, which constrain the total cross section to lie within a factor of a few of the SM prediction up to $E \sim 10\,\mathrm{PeV}$~\cite{IceCube:2020rnc}.  

We note that the magnetic-moment-dipole interaction implies also the upscattering process $\nu \mathrm{e} \to N  \mathrm{e}$. However, for an electron target, the available center-of-mass energy is smaller than that for a nucleus target and the $\nu \mathrm{e}$ process can only contribute at higher $E$ values. The contribution is subdominant in the parameter ranges considered in this study.

\section{Exposure to ultrahigh energy neutrinos}
\label{sec:exposure}

In this section, we evaluate the  exposure of the Observatory to UHE$\nu$s by accounting for the magnetic-induced transitions. As will become evident, the down-going and Earth-skimming detection modes respond in markedly different ways. In the down-going channel, the upscattering process $\nu\mathrm{n}\rightarrow NX$, followed by the prompt decay of the heavy neutrino, produces final states that closely resemble those of SM neutrino interactions. By contrast, in the Earth-skimming channel, the heavy neutrino decays back into a $\nu_\uptau$, so that each upscattering acts as a neutral-current-like interaction in which the $\nu_\uptau$ retains on average a fraction $50\%$ of its energy. Since the BSM interaction length can be much shorter than the CC one (the latter amounting to $\sim\!600$\,km of rock at 1\,EeV) while the rock chords traversed range from $\sim\!200$ to $\sim\!1100$\,km in the angular range of interest, several such upscatterings may occur before a CC interaction, until the $\nu_\uptau$ energy falls below the kinematic threshold $m_N^2/2m_\mathrm{n}$. For most of the $m_N$ range considered, this threshold lies at or below the energies ($\gtrsim0.2$\,EeV) at which emerging $\uptau$ leptons can be detected.

\subsection{Down-going channel}
\label{sec:dg}

In the down-going channel, the exposure can be expressed as
\begin{multline}
    \label{eqn:exponu-dg}
    \mathcal{E}^{\mathrm{DG}}_\nu(E;m_N,\mu_{\nu N})=
    m_\mathrm{n}^{-1}\sigma_{\nu\rm n}(E;m_N,\mu_{\nu N})\\
    \times \iiiint \dif\Omega\dif X\dif A\dif t\cos{\theta}\,\epsilon_\mathrm{DG}(E,\Omega,X,\mathbf{x},t;m_N),
\end{multline}
where the integrations extend over the solid angle $\Omega=(\theta,\varphi)$ with the zenith angle restricted to $75^\circ \le \theta < 90^\circ$ and the azimuth $\varphi$ ranging from 0 to $360^\circ$~\cite{PierreAuger:2019ens}, the slant depth $X$ at which the neutrino interaction occurs, the detector area $A$ scanned over position $\mathbf{x}$, and observation time $t$. 

The SM detection efficiency, $\epsilon_{\mathrm{DG}}$, accounts for the sum over neutrino flavors and interaction channels~\cite{PierreAuger:2011cpc,PierreAuger:2015ihf,PierreAuger:2019ens}. All neutrino interactions, regardless of flavor, share the feature of producing secondary particles (in particular the ``debris'') that initiate an extensive air shower. CC interactions of electron neutrinos further enhance the air-shower development due to the additional energy deposited by the outgoing electron (same in the case of CC interactions of tau neutrinos in the event of a decay of the outgoing tau lepton). 

We follow the same procedure to calculate  $\epsilon_\mathrm{DG}$ in the presence of $\nu$--$N$ transitions. The energy carried out by the debris is primarily determined by the inelasticity of the reaction that depends on $m_N$. The air shower is further fed by the decay products of the heavy neutrino, which disintegrates promptly into $N \to \nu \gamma$ and  $N \to \nu Z$, with branching fractions given by, in the limit $m_N\gg m_Z$, $\text{BR}(N \to \nu \gamma)  = \cos^2 \theta_W   \approx 77\%$ and $\text{BR}(N \to \nu Z) = \sin^2\theta_W  \approx 23\%$, 
where $\theta_W$ is the weak mixing angle~\cite{Zhang:2023nxy,Ovchynnikov:2023wgg,Brdar:2025iua}.\footnote{This differs sharply from the standard Yukawa mixing portal where, by virtue of the Goldstone boson equivalence theorem~\cite{Cornwall:1974km}, multi-TeV sterile neutrino decays exhibit
flavor-blind, asymptotically fixed branching ratios: $\text{BR}(N\to W^{\pm }\ell ^{\mp })\simeq 50\%$, $\text{BR}(N\to Z\nu )\simeq 25\%$, and $\text{BR}(N\to h\nu )\simeq 25\%$~\cite{Anchordoqui:2026ret}.} Decays induced by an active--sterile mixing $U$ are negligible: their width relative to the dipole one, $\Gamma_\mathrm{mix}/\Gamma_\mathrm{dip}\sim G_F|U|^2/(\sqrt{2}\,\mu_{\nu N}^2)$, with $G_F$ the Fermi constant, remains below $10^{-3}$ in the excluded region given the electroweak-precision bound $|U|^2\lesssim10^{-3}$ for TeV-scale masses~\cite{Blennow:2023mqx}. For $m_N\gtrsim m_Z$, electroweak fragmentation of the decay products steadily enters into play and is included using HDMSpectra~\cite{Bauer:2020jay}. The dependence of $\epsilon_{\mathrm{DG}}$ on $m_N$ for a reference value of $\mu_{\nu N}$ is thus traced through air showers featuring the characteristics of the magnetic-moment-dipole interaction.


\begin{figure}[!ht]
\centering
\includegraphics[width=0.5\textwidth]{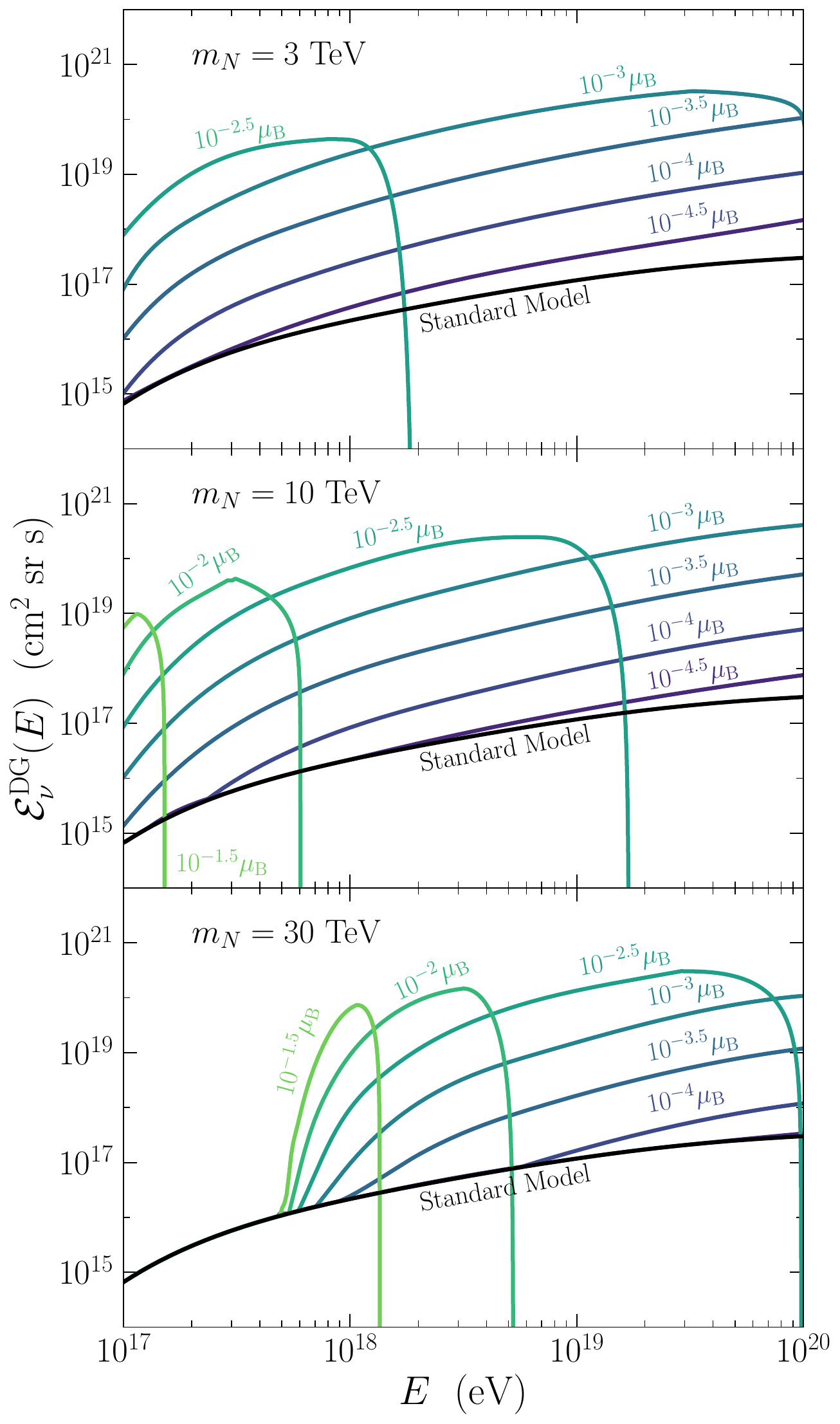}
\caption{Exposure to down-going neutrinos for $m_N=3\,$TeV (top), $10\,$TeV (middle), and $30\,$TeV (bottom). Black curves: SM cross section alone. Colored curves: including the upscattering channel, with $\mu_{\nu N}$ progressing from $10^{-4.5}\mu_\mathrm{B}$ to $10^{-1.5}\mu_\mathrm{B}$, color-coded from dark blue to light green.}
\label{fig:expoDG}
\end{figure}

The dependence of $\mathcal{E}^{ \mathrm{DG}}_{\nu}$ on $\mu_{\nu N}$ follows that of $\sigma_{\nu \mathrm{n}}$, on the condition that $\sigma_{\nu \mathrm{n}}$ remains smaller than cross sections typical of strong or electromagnetic interactions. The term $m_\mathrm{n}^{-1}\,\sigma_{\nu \mathrm{n}}(E; m_N, \mu_{\nu N})\,\dif X$ in Eq.~\eqref{eqn:exponu-dg} represents the average number of interactions of a neutrino  within an infinitesimal column depth $\dif X$. When integrated over the entire slant depth traversed from the top of the atmosphere to the ground level $X_\mathrm{ground}(\theta)$, this number must remain below unity to ensure that the interaction can occur with almost equal probability at any point between the top of the atmosphere and the ground. This condition is required because, for sufficiently large cross sections, neutrinos would interact too high in the atmosphere to be distinguishable from the cosmic-ray background. We therefore impose conservatively that the efficiency vanishes in Eq.~\eqref{eqn:exponu-dg} whenever
\begin{equation}
    \label{eqn:sigma_mn_dX}
       m_\mathrm{n}^{-1}\sigma_{\nu \mathrm{n}}(E; m_N, \mu_{\nu N})X_\mathrm{ground}(\theta)
    \geq 1.
\end{equation}

The resulting exposure is shown in Fig.~\ref{fig:expoDG} for $m_N = 3$, 10, and 30~TeV (top, middle, and bottom panels, respectively) and various values of the transition magnetic moment $\mu_{\nu N}$. The black lines represent the benchmark expectation obtained under the assumption of SM neutrino interactions. The BSM contribution sets in above the kinematic threshold $m_N^2/2m_\mathrm{n}\simeq 5$~PeV, 53~PeV, and 0.48~EeV, respectively; for a given $\mu_{\nu N}$, the excess relative to the SM prediction therefore decreases with increasing $m_N$, and for $m_N=30$~TeV the onset of the upscattering is directly visible within the energy range shown. The excess becomes noticeable once $\mu_{\nu N}\gtrsim 10^{-4.5}$--$10^{-4}\,\mu_\mathrm{B}$, depending on $m_N$, and scales as $\mu_{\nu N}^2$ up to $\mu_{\nu N}\simeq 10^{-3}\,\mu_\mathrm{B}$. For higher values, the exposure drops to zero above a cutoff energy. This energy cutoff decreases with increasing $\mu_{\nu N}$ within each panel of Fig.~\ref{fig:expoDG}, and increases with $m_N$ from one panel to the next. This behavior originates from the condition imposed to ensure the validity of the down-going detection efficiency $\epsilon_{\mathrm{DG}}$ (cf. Eq.~\eqref{eqn:sigma_mn_dX}). For $\mu_{\nu N}\gtrsim 10^{-1}\,\mu_\mathrm{B}$, the exposure vanishes over the whole energy range shown for the values $m_N=3$ and 10~TeV, and survives only below the kinematic threshold for $m_N=30$~TeV.
 
Several sources of systematic effects affect the exposure calculation~\cite{PierreAuger:2009dvq}. The impact of the simulation framework was assessed through dedicated comparisons between different implementations: HERWIG~\cite{Corcella:2000bw} and PYTHIA~\cite{Sjostrand:2006za} for neutrino interactions, AIRES~\cite{Sciutto:1999xz} and CORSIKA~\cite{Heck:1998vt} for air-shower simulations, SIBYLL~\cite{Fletcher:1994bd} and QGSJET~\cite{Kalmykov:1997te}  for hadronic interactions, and thinning levels of $10^{-7}$ and $10^{-6}$. Thinning is a statistical sampling technique used to make the simulation of air showers tractable at ultrahigh energies: below a given fraction (the thinning level) of the primary energy, only a subset of the secondary particles is followed, with weights assigned to conserve energy~\cite{Hillas:1997tf}. These studies lead to an estimated contribution of $\pm 4\%$. In addition, the uncertainty in the neutrino-nucleon cross section arising from PDFs, investigated in Ref.~\cite{Cooper-Sarkar:2007zsa}, amounts to 5\%--9\%. This estimate includes both experimental limitations of the PDF fits derived from fixed-target data and theoretical effects associated with the treatment of heavy-quark masses in PDF evolution, assuming DGLAP evolution over the relevant kinematic range extending down to $x\sim10^{-12}$.

\subsection{Earth-skimming channel}
\label{sec:es}

In the Earth-skimming channel, the exposure is obtained by integrating the detection efficiency over the solid angle $\Omega=(\theta,\varphi)$ restricted to $90^\circ \le \theta < 95^\circ$, the detector area $A$ scanned over position $\mathbf{x}$ and observation time $t$, 
\begin{multline}
    \label{eqn:exposureN}
    \mathcal{E}^{\mathrm{ES}}_{\nu_\uptau}(E;m_N,\mu_{\nu N})=\\
    \iiint\dif\Omega\,\dif A\,\dif t\, \left|\cos\theta\right|\,\epsilon_\mathrm{ES}(E,\Omega,\mathbf{x},t;m_N,\mu_{\nu N}).
\end{multline}
The detection efficiency for $\nu_\uptau$ is obtained by folding the distribution of $\uptau$ leptons emerging from the Earth with the efficiency for detecting the air showers initiated by their decays in the atmosphere above the array,
\begin{multline}
    \label{eqn:effN}
    \epsilon_\mathrm{ES}(E,\Omega,\mathbf{x},t)=\iint\dif\Omega_\uptau\dif E_\uptau\,\frac{\dif^2n_\uptau(E_\uptau,\Omega_\uptau;E,\Omega)}{\dif\Omega_\uptau\dif E_\uptau}\\
    \times \int\dif s\,\frac{\mathrm{e}^{-s/\lambda(E_\uptau)}}{\lambda(E_\uptau)}\epsilon_\uptau(E_\uptau,\Omega_\uptau,s\left|\cos\theta_\uptau\right|,\mathbf{x},t).
\end{multline}
Here, $\dif^2n_\uptau/\dif\Omega_\uptau\dif E_\uptau$ denotes the number of $\uptau$ leptons emerging from the surface of the Earth with energy $E_\uptau$ in the direction $\Omega_\uptau$, per incident $\nu_\uptau$ with energy $E$ and direction $\Omega$. At ultrahigh energies, the $\uptau$ leptons are emitted along the direction of the incident neutrino, so that $\Omega_\uptau\simeq\Omega$. The last integral accounts for the probability that the $\uptau$ decays after a distance $s$ along its trajectory, i.e., at an altitude $s\left|\cos\theta_\uptau\right|$, given its decay length $\lambda\simeq 49\,(E_\uptau/\mathrm{EeV})$\,km. This distribution is estimated by Monte Carlo simulations, in which $n_{\nu_\uptau}$ tau neutrinos are propagated through the Earth along directions $\Omega$ drawn at random within $\theta\in[90^\circ,95^\circ]$ and $\varphi\in[0^\circ,360^\circ]$, yielding $n_\uptau$ emerging $\uptau$ leptons:
\begin{equation}
    \label{eqn:dNdtau}
    \frac{\dif^2n_\uptau}{\dif \Omega_\uptau\dif E_\uptau}=\frac{1}{n_{\nu_\uptau}}\sum_{i=1}^{n_\uptau(E,\Omega)}\delta(E_\uptau,E_{\uptau i})\delta(\Omega_\uptau,\Omega_{\uptau i}).
\end{equation}
In the simulation, the Earth density is modeled as in~\cite{EarthGeophysics}. A step-by-step method is used: at each step, the probabilities of the different interactions and of the decay for the $\uptau$ are evaluated as functions of the energy. Both SM and the BSM interactions are accounted for. The BSM interaction closely resembles an SM neutral current (NC) interaction because the produced heavy neutrino decays promptly with a $\nu_\uptau$ in final states, leaving mainly a modified energy partition determined by the interaction inelasticity. Energy losses of the $\uptau$ include Bremsstrahlung, pair production and DIS. The energy loss model is of the form $\dif E_\uptau/\dif x\simeq -b(E_\uptau)E_\uptau$, with $b(E_\uptau)=(1.2+0.16\,\ln (E_\uptau/{10\,\mathrm{EeV}}))\,10^{-6}$\,g$^{-1}$\,cm$^2$~\cite{Dutta:2000hh,Armesto:2007tg}. 

Overall, the final expression of $\mathcal{E}^\mathrm{ES}_{\nu_\uptau}$ makes use of the Monte Carlo estimation of the detection efficiency to up-going showers integrated over time $\Delta T$ and effective surface $S$,
\begin{multline}
    \label{eqn:exposureNbis}
    \mathcal{E}_\mathrm{ES}(E;m_N,\mu_{\nu N})=\frac{S\,\Delta T}{n_{\nu_\uptau}}\int\dif\Omega\,\left|\cos\theta\right|\\
    \times\sum_{i}^{n_\uptau} \int\dif s\,\frac{\mathrm{e}^{-s/\lambda(E_{\uptau i})}}{\lambda(E_{\uptau i})}\langle\epsilon_\uptau(E_{\uptau i},\Omega_{\uptau i},s\left|\cos\theta_i\right|)\rangle_{\mathbf{x},t},
\end{multline}
with $n_\uptau\equiv n_\uptau(E,\Omega;m_N,\mu_{\nu N})$. Here, the factor $S\,\Delta T$ results from the integration over detector area and observation time already present in Eq.~\eqref{eqn:exposureN}, the efficiency being averaged over the position and time, $\langle\cdots\rangle_{\mathbf{x},t}$. The last integration over solid angle yields a global factor $\pi\sin^2 5^\circ$ given the way the Monte-Carlo is conducted to estimate $\dif^2n_\uptau/\dif\Omega_\uptau\dif E_\uptau$.

\begin{figure}[t]
\centering
\includegraphics[width=0.5\textwidth]{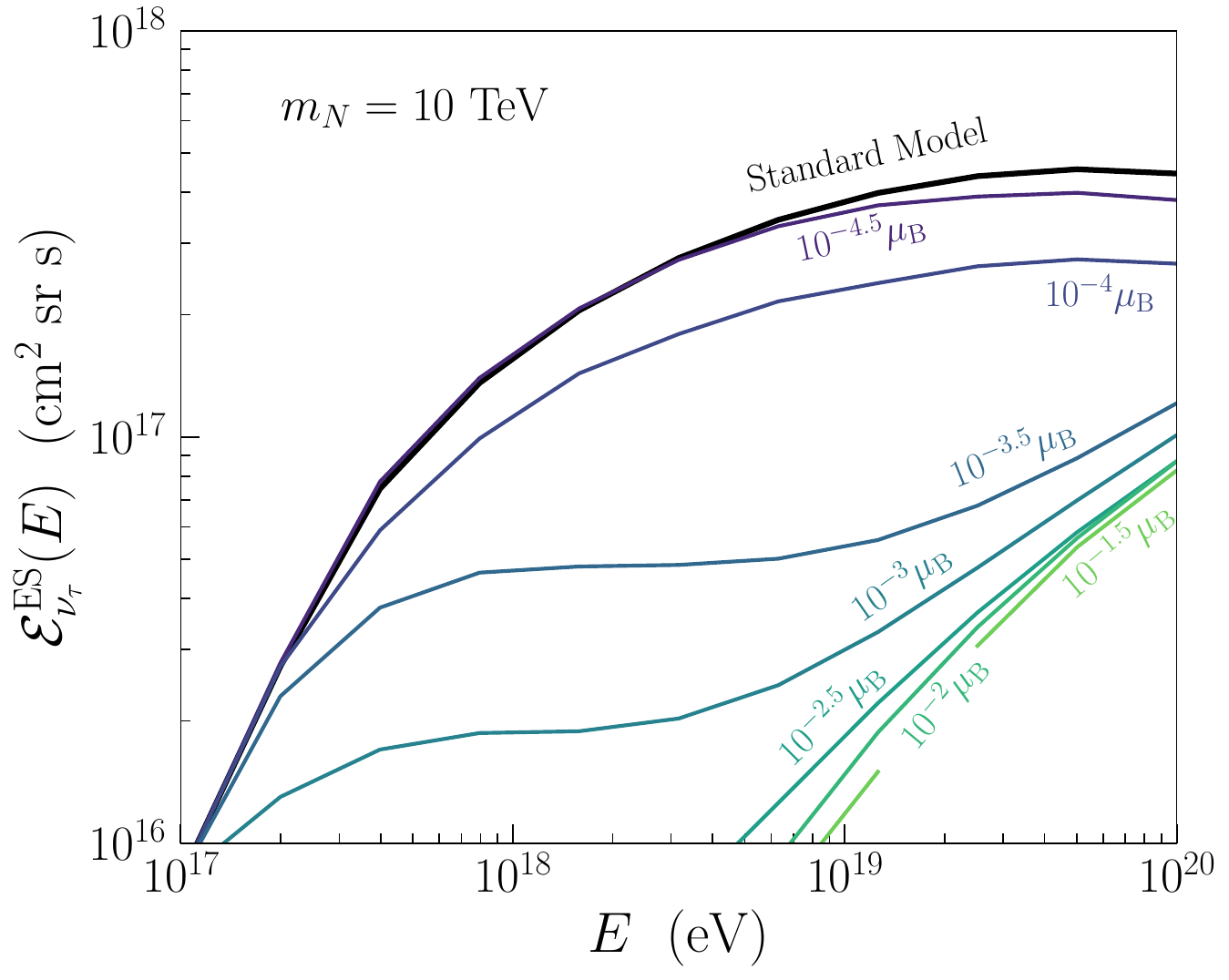}
\caption{Exposure to Earth-skimming $\uptau$-neutrinos.  Black: SM. Color lines: BSM with $m_N=10\,$TeV, using the same color code as in Fig.~\ref{fig:expoDG} for $\mu_{\nu N}$ values.}
\label{fig:expoES}
\end{figure}

The resulting exposure for the Earth-skimming channel is shown in Fig.~\ref{fig:expoES} as a function of the neutrino energy, assuming a sterile neutrino mass of $m_N = 10~\mathrm{TeV}$. As expected, the exposure decreases with increasing values of $\mu_{\nu N}$. This behavior results from the growing importance of  NC interactions over CC ones: the BSM contribution effectively enhances the NC channel, thereby increasing the probability that the neutrino loses energy without producing the charged lepton required to initiate an observable air shower. This trend exemplifies contrasting responses in the down-going and Earth-skimming detection channels to the BSM scenario.  

\section{Results}
\label{sec:constraints}

The limits are derived from the down-going event selection of Refs.~\cite{PierreAuger:2011cpc,PierreAuger:2019ens}, applied to data collected from 1 January 2004 to  31 December 2021, in which no neutrino candidate was identified. Only the down-going exposure $\mathcal{E}^\mathrm{DG}_\nu$ enters $n_\mathrm{exp}$: since the Earth-skimming exposure is reduced by the BSM interaction (Sec.~\ref{sec:es}), omitting it is conservative. The fluxes $\Phi_\nu$ are single-flavor fluxes, and flavor equipartition at Earth is assumed, consistent with the sum over flavors entering $\epsilon_\mathrm{DG}$. The transition moment is taken to be flavor-diagonal and identical for the three flavors, so that a heavy neutrino produced by a $\nu_\alpha$ decays back into a $\nu_\alpha$.  The systematic uncertainties on the exposure discussed in Sec.~\ref{sec:dg} (at the 10\% level) are not propagated into the limits: since $n_\mathrm{exp}\propto\mu_{\nu N}^2$ over most of the excluded region, they translate into a $\sim$5\% shift of the bounds on $\mu_{\nu N}$, negligible compared to the spread induced by the flux scenarios.

We select expectations of the neutrino flux by requiring that the models reproduce the mass-discriminated cosmic-ray spectra inferred from the Observatory data above $5~\mathrm{EeV}$. A conservative estimate stems from the production \textit{en route} alone. For various source evolutions, there is convergence in the literature for a (single-flavor) flux scaling as $E^2\Phi_\nu\simeq 2\times 10^{-10}~\mathrm{GeV\,cm^{-2}\,sr^{-1}\,s^{-1}}$ at $100\,$PeV and decreasing in the energy range of interest for this study down to $E^2 \Phi_\nu \simeq 10^{-12}~\mathrm{GeV~cm^{-2}~sr^{-1}~s^{-1}}$ at $\simeq 1~\mathrm{EeV}$; specifically, we follow the GRB-like source evolution of Ref.~\cite{AlvesBatista:2018zui}. When considering specifics of in-source production such as tidal disruption events~\cite{Biehl:2017hnb}, low-luminosity gamma-ray bursts~\cite{Boncioli:2018lrv} or starburst-galaxy environments~\cite{Muzio:2022bak}, the expected fluxes reach $E^2\Phi_\nu\simeq 2\times10^{-9}~\mathrm{GeV\,cm^{-2}\,sr^{-1}\,s^{-1}}$ at the peak energy $100\,$PeV; we follow here Ref.~\cite{Muzio:2022bak}. Alternatively, nonminimal UHECR models invoke the interplay of two source populations, one of which accelerates a subdominant component of protons up to, or even above, $100~\mathrm{EeV}$~\cite{Rodrigues:2020pli,PierreAuger:2022atd,Muzio:2023skc,Ehlert:2023btz}. Even a small fraction of protons then significantly enhances the neutrino flux to an almost constant level $E^2\Phi_\nu\simeq 2\times10^{-9}~\mathrm{GeV\,cm^{-2}\,sr^{-1}\,s^{-1}}$ between $100$\,PeV and $30~$EeV. One of the major goals of the upgraded Observatory~\cite{PierreAuger:2016qzd} is to improve the sensitivity to a proton component extending to the highest energies, with the goal of constraining or uncovering proton fractions even below the 10\% level. These measurements will considerably tighten the allowed range of neutrino fluxes in the near future; as a result, the exclusion regions will be defined more sharply.

We denote these scenarios by ``$\Phi_\nu$-low'',``$\Phi_\nu$-high'', and ``$\Phi_\nu$-with UHE p'', corresponding respectively to minimal, optimistic-minimal, and nonminimal flux predictions.  To derive constraints in the plane $(m_N,\mu_{\nu N})$, we calculate the expected number of events as
\begin{equation}
    n_\mathrm{exp}(m_N,\mu_{\nu N})=\int\dif E\,\Phi_\nu(E)\mathcal{E}_\nu^\mathrm{DG}(E;m_N,\mu_{\nu N}).
\end{equation}
The regions of $(m_N,\mu_{\nu N})$ values that lead to $n_\mathrm{exp}>2.44$, appropriate in the absence of neutrino candidates and in background-free regime~\cite{Feldman:1997qc}, can be excluded with 90\% C.L. 

\begin{figure}[t]
\centering
\includegraphics[width=0.5\textwidth]{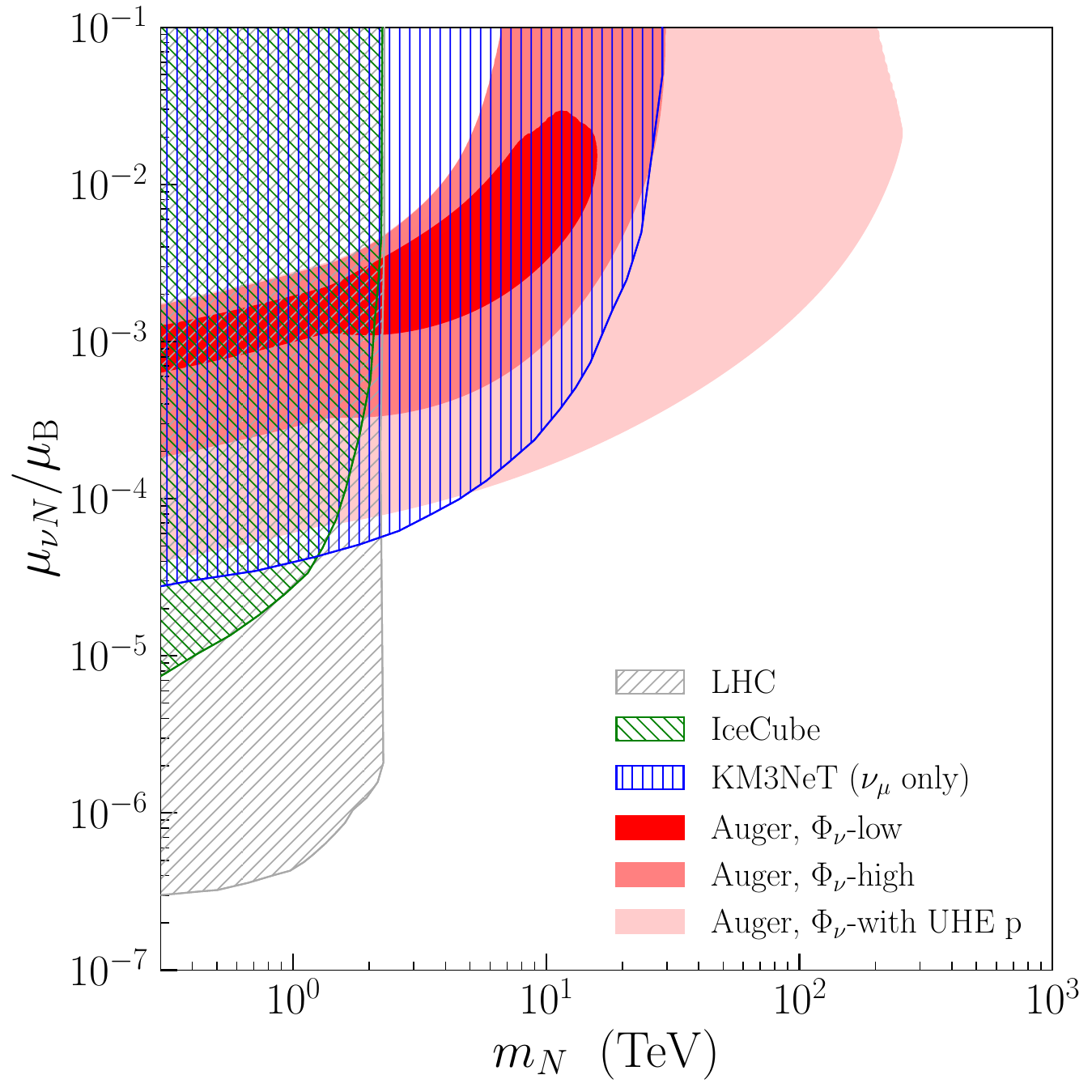}
\caption{Excluded regions in the $(m_N, \mu_{\nu N})$ plane derived from the non-observation of UHE neutrinos at the Observatory. The dark red, medium red, and light red regions correspond respectively to the ``$\Phi_\nu$-low'', ``$\Phi_\nu$-high'', and ``$\Phi_\nu$-with-UHE-p'' neutrino flux scenarios. For comparison, existing bounds from IceCube~\cite{Huang:2022pce}, LHC~\cite{Magill:2018jla} and the KM3-230213A event~\cite{Munoz-Ovalle:2025riq} are also shown.}
\label{fig:limits}
\end{figure}

The resulting constraints are displayed in Fig.~\ref{fig:limits}, with dark red indicating the most conservative ``$\Phi_\nu$-low'' scenario, red corresponding to the ``$\Phi_\nu$-high'' scenario, and light red representing the alternative ``$\Phi_\nu$-with-UHE-p'' scenario. As expected, the limits become more stringent in scenarios with higher neutrino fluxes. Notably, the constraints extend into the region $m_N > 1~\mathrm{TeV}$, which is accessible only to UHE neutrino observatories. The lower ends of the contours are shaped primarily by the fall-off of the incoming fluxes, whereas the upper ends are determined by the requirement encapsulated in Eq.~\eqref{eqn:sigma_mn_dX}, which ensures the applicability of the detection efficiency.  

For comparison, constraints inferred from recasts of IceCube, LHC and KM3NeT data are shown. The IceCube excluded region follows from the absence of significant attenuation of the observed high-energy neutrino flux~\cite{Huang:2022pce}. The LHC excluded region, on the other hand, is derived from ATLAS and CMS searches for final states with a single photon plus missing-energy signatures produced through $q\bar q\to N\nu$ followed by $N\to\gamma\nu$~\cite{Magill:2018jla}. Finally, the KM3NeT excluded region follows from the observation of KM3-230213A and the requirement that dipole-induced upscattering does not excessively degrade the neutrino energy during propagation through Earth. Among the constraints derived in Ref.~\cite{Munoz-Ovalle:2025riq}, we show the one obtained for the nominal energy and arrival direction of KM3-230213A. By contrast to ours, these constraints are unbounded from above in the $\mu_{\nu N}$ direction in Fig.~\ref{fig:limits}. Yet, our constraints probe a complementary region of parameter space, and extend the coverage to electron- and tau-neutrino flavors compared to those from KM3-230213A that pertain to the muon-flavor only. 

We note that the dipole interaction is treated as an effective operator (Appendix~\ref{sec:app1}). For the excluded values, $\mu_{\nu N}^{-1}$ lies below $m_N$ (i.e., $\mu_{\nu N}\gtrsim 3.4\times10^{-6}\,\mu_\mathrm{B}\,(1\,\mathrm{TeV}/m_N)$) and below the typical momentum transfers. An ultraviolet completion would then involve new states lighter than the probed $Q$, possibly modifying the $Q^2$ dependence of the interaction and the decay of $N$. The constraints of Fig.~\ref{fig:limits} must thus be understood as bounds on the effective coupling, a caveat that applies equally to the IceCube, LHC, and KM3NeT recasts.

\section{Conclusions}
\label{sec:conclusions}

We have investigated the sensitivity of the Observatory to magnetic-moment–induced transitions between active and sterile neutrino states. Such transitions, characterized by a coupling $\mu_{\nu N}$, can enhance the neutrino--nucleon and neutrino--electron cross sections above a kinematic threshold determined by the sterile neutrino mass $m_N$, leading to potentially observable deviations from SM expectations. Interestingly, we find that, focusing on sterile neutrinos with masses in the $1\,$TeV--$100\,$TeV range, magnetic-moment–induced interactions enhance the down-going event rate, while the Earth-skimming channel can experience a relative suppression, providing a distinctive diagnostic to disentangle flux enhancements from cross-section modifications.

To quantify the implications for BSM physics, we considered three benchmark neutrino flux scenarios, encompassing minimal and nonminimal assumptions on UHECR sources and composition. Using the absence of observed UHE neutrino events, we derived 90\% C.L. flux-dependent constraints in the $(m_N, \mu_{\nu N})$ plane. The resulting limits probe sterile neutrino masses above 1~TeV, extending existing constraints from recasts of IceCube and LHC data. Moreover, because they are flavor-independent, these limits provide complementary coverage of the electron- and tau-neutrino flavors, alongside the partially overlapping parameter space explored in the muon-neutrino channel through the KM3-230213A event. Overall, this work demonstrates that the Observatory offers sensitivity to BSM interactions at energy scales beyond the reach of laboratory experiments.


\begin{acknowledgements}
\begin{sloppypar}
The successful installation, commissioning, and operation of the Pierre
Auger Observatory would not have been possible without the strong
commitment and effort from the technical and administrative staff in
Malarg\"ue. We are very grateful to the following agencies and
organizations for financial support:
\end{sloppypar}

\begin{sloppypar}
Argentina -- Comisi\'on Nacional de Energ\'\i{}a At\'omica; Agencia Nacional de
Promoci\'on Cient\'\i{}fica y Tecnol\'ogica (ANPCyT); Consejo Nacional de
Investigaciones Cient\'\i{}ficas y T\'ecnicas (CONICET); Gobierno de la
Provincia de Mendoza; Municipalidad de Malarg\"ue; NDM Holdings and Valle
Las Le\~nas; in gratitude for their continuing cooperation over land
access; Australia -- the Australian Research Council; Belgium -- Fonds
de la Recherche Scientifique (FNRS); Research Foundation Flanders (FWO),
Marie Curie Action of the European Union Grant No.~101107047; Brazil --
Minist\'erio da Ci\^encia, Tecnologia e Inova\c{c}\~ao (MCTI); Czech Republic --
GACR 24-13049S, CAS LQ100102401, MEYS LM2023032,
CZ.02.1.01/0.0/0.0/16{\textunderscore}013/0001402, CZ.02.1.01/0.0/0.0/18{\textunderscore}046/0016010
and CZ.02.1.01/0.0/0.0/17{\textunderscore}049/0008422 and
CZ.02.01.01/00/22{\textunderscore}008/0004632; France -- Centre de Calcul IN2P3/CNRS;
Centre National de la Recherche Scientifique (CNRS); Institut National
de Physique Nucl\'eaire et de Physique des Particules (IN2P3/CNRS);
Germany -- Bundesministerium f\"ur Forschung, Technologie und Raumfahrt
(BMFTR); Deutsche Forschungsgemeinschaft (DFG); Ministerium f\"ur Finanzen
Baden-W\"urttemberg; Helmholtz Alliance for Astroparticle Physics (HAP);
Hermann von Helmholtz-Gemeinschaft Deutscher Forschungszentren e.V.;
Ministerium f\"ur Kultur und Wissenschaft des Landes Nordrhein-Westfalen;
Ministerium f\"ur Wissenschaft, Forschung und Kunst des Landes
Baden-W\"urttemberg; Italy -- Istituto Nazionale di Fisica Nucleare
(INFN); Istituto Nazionale di Astrofisica (INAF); Ministero
dell'Universit\`a e della Ricerca (MUR); CETEMPS Center of Excellence;
Ministero degli Affari Esteri (MAE), ICSC Centro Nazionale di Ricerca in
High Performance Computing, Big Data and Quantum Computing, funded by
European Union NextGenerationEU, reference code CN{\textunderscore}00000013; M\'exico --
Consejo Nacional de Ciencia y Tecnolog\'\i{}a (CONACYT-SECHTI)
No.~CB-A1-S-46703, Universidad Nacional Aut\'onoma de M\'exico (UNAM)
PAPIIT-IN114924; Benem\'erita Universidad Aut\'onoma de Puebla (BUAP), VIEP
and Laboratorio Nacional de Superc\'omputo del Sureste de M\'exico (LNS);
and Benem\'erita Universidad Aut\'onoma de Chiapas (UNACH); The Netherlands
-- Ministry of Education, Culture and Science; Netherlands Organisation
for Scientific Research (NWO); Dutch national e-infrastructure with the
support of SURF Cooperative; Poland -- Ministry of Science and Higher
Education, grant No.~2026/WK/05; National Science Centre, grant
No.~2022/45/B/ST9/02163; Portugal -- Portuguese national funds and FEDER
funds within Programa Operacional Factores de Competitividade through
Funda\c{c}\~ao para a Ci\^encia e a Tecnologia (COMPETE); Romania -- Ministry of
Education and Research, contract no.~30N/2023 under Romanian National
Core Program LAPLAS VII, and grant no.~PN 23 21 01 02; Slovenia --
Slovenian Research and Innovation Agency, grants P1-0031, I0-0033; Spain
-- Ministerio de Ciencia, Innovaci\'on y Universidades/Agencia Estatal de
Investigaci\'on MICIU/AEI /10.13039/501100011033 (PID2022-140510NB-I00,
PCI2023-145952-2, CNS2024-154676, and Mar\'\i{}a de Maeztu CEX2023-001318-M),
Xunta de Galicia (CIGUS Network of Research Centers, Consolidaci\'on
ED431C-2025/11 and ED431F-2022/15) and European Union ERDF; USA --
Department of Energy, Contracts No.~DE-AC02-07CH11359,
No.~DE-FR02-04ER41300, No.~DE-FG02-99ER41107 and No.~DE-SC0011689;
National Science Foundation, Grant No.~0450696, and NSF-2013199; The
Grainger Foundation; Astrophysics Centre for Multi-messenger studies in
Europe (ACME) EU Grant No 101131928; and UNESCO.
\end{sloppypar}

\end{acknowledgements}

\appendix 

\section{Active-to-sterile neutrino transition magnetic moment}
\label{sec:app1}


In this work, we consider an effective active-to-sterile neutrino transition magnetic moment interaction. Following the benchmark scenarios commonly considered in the literature~\cite{Magill:2018jla,Zhang:2023nxy,Barducci:2024kig,Brdar:2025iua}, we assume that the sterile neutrino possesses transition dipole couplings only to the hypercharge gauge boson. After electroweak symmetry breaking, the dipole portal interaction  to the neutral electroweak gauge bosons takes the form~\cite{Magill:2018jla}
\begin{multline}
    \label{eqn:app2}
    \mathcal{L}_{\mu_{\nu N}}
    =\mu_{\nu N}\,
    \overline{\nu}_\mathrm{L}\sigma_{\mu\nu}N_\mathrm{R}F^{\mu\nu}\\
    -\mu_{\nu N}\tan\theta_\mathrm{W}\,
    \overline{\nu}_\mathrm{L}\sigma_{\mu\nu}N_\mathrm{R}Z^{\mu\nu}
    +\mathrm{h.c.},
\end{multline}
where $\nu$ and $N$ denote the active and sterile neutrino fields, respectively. Here, $F^{\mu\nu}$ and $Z^{\mu\nu}$ are the field strength tensors associated with the photon and the $Z$ boson, respectively, $\theta_\mathrm{W}$ is the weak mixing angle, and
$\sigma_{\mu\nu}=\frac{i}{2}(\gamma_\mu\gamma_\nu-\gamma_\nu\gamma_\mu)$. Note that we absorb the $\cos\theta_\mathrm{W}$ factor associated with the hypercharge gauge boson coupling before electroweak symmetry breaking, so that our definition of $\mu_{\nu N}$ is consistent with the existing limits in the literature, which generally assume a photon-only coupling.

The transition magnetic moment interaction is treated as an effective operator arising after integrating out heavy states of an underlying ultraviolet completion. The effective description is therefore valid only when the characteristic momentum transfer $Q$ involved in the scattering process remains below the mass scale of the particles generating the dipole interaction. Accordingly, it is commonly assumed that the masses of the states in the ultraviolet completion are larger than the typical momentum transfers relevant for the processes considered~\cite{Huang:2022pce}.

The differential cross section in the DIS regime is inferred from 
\begin{equation}
    \frac{\dif^2\sigma}{\dif x\dif y}=\sum_i\mathrm{q}_i(x,Q^2)\frac{1}{16\pi\hat{s}}\overline{|\mathcal{M}|^2},
\end{equation}
with $\mathcal{M}$ resulting from the sum of the photon contribution on the one hand,
\begin{equation}
    \mathcal{M_\gamma}=\frac{eQ_i\mu_{\nu N}}{t}(\overline{u}_N\sigma_{\mu\nu}q^\nu u_{\nu_\LL})(\overline{u}_\mathrm{q}\gamma^\mu u_\mathrm{q}),
\end{equation}
and of the $Z$ boson one on the other hand,
\begin{multline}
    \mathcal{M}_Z=\mathcal{M}_Z^\mathrm{V}+\mathcal{M}_Z^\mathrm{A}=\frac{g\mu_{\nu N}\tan\theta_\mathrm{W}}{2\cos\theta_{\mathrm{W}}(t-m_Z^2)}\\
    \times (\overline{u}_N\sigma_{\mu\nu}q^\nu u_{\nu_\LL})\left(\overline{u}_\mathrm{q}\gamma^\mu (g_i^\mathrm{V}-g_i^\mathrm{A}\gamma^5)u_\mathrm{q}\right),
\end{multline}
where $u_i$ denotes the Dirac spinor wavefunction for incoming and outgoing particles. The averaged squared-matrix element consists then of four nonzero terms,
\begin{equation}
    \overline{|\mathcal{M}|^2}=\overline{|\mathcal{M}_\gamma|^2}+\overline{|\mathcal{M}_Z^\mathrm{V}|^2}+\overline{|\mathcal{M}_Z^\mathrm{A}|^2}+2\mathrm{Re}\left(\mathcal{M}_\gamma(\mathcal{M}_Z^\mathrm{V})^\star\right),
\end{equation}
given that $2\mathrm{Re}\left(\mathcal{M}_\gamma(\mathcal{M}_Z^\mathrm{A})^\star\right)=2\mathrm{Re}\left(\mathcal{M}_Z^\mathrm{V}(\mathcal{M}_Z^\mathrm{A})^\star\right)=0$. The contractions and summations lead to Eq.~\eqref{eqn:dsig_dxdy}, which reproduces the results obtained in the literature for a photon-mediated interaction~\cite{Magill:2018jla,Miranda:2021kre,Huang:2022pce}, while consistently extending them to include $Z$-boson exchange. In particular, the inclusion of $Z$ exchange and its interference with the photon amplitude increases the cross section by approximately $15\%$ relative to the photon-only case, thereby providing a more complete description of the process.



\bibliographystyle{apsrev4-2} 
\bibliography{biblio}

\end{document}

%% file: revtex_authorlist.tex

\affiliation{Centro At\'omico Bariloche and Instituto Balseiro (CNEA-UNCuyo-CONICET), San Carlos de Bariloche, Argentina}
\affiliation{Departamento de F\'\i{}sica and Departamento de Ciencias de la Atm\'osfera y los Oc\'eanos, FCEyN, Universidad de Buenos Aires and CONICET, Buenos Aires, Argentina}
\affiliation{IFLP, Universidad Nacional de La Plata and CONICET, La Plata, Argentina}
\affiliation{Instituto de Astronom\'\i{}a y F\'\i{}sica del Espacio (IAFE, CONICET-UBA), Buenos Aires, Argentina}
\affiliation{Instituto de F\'\i{}sica de Rosario (IFIR) -- CONICET/U.N.R.\ and Facultad de Ciencias Bioqu\'\i{}micas y Farmac\'euticas U.N.R., Rosario, Argentina}
\affiliation{Instituto de Tecnolog\'\i{}as en Detecci\'on y Astropart\'\i{}culas (CNEA, CONICET, UNSAM), and Universidad Tecnol\'ogica Nacional -- Facultad Regional Mendoza (CONICET/CNEA), Mendoza, Argentina}
\affiliation{Instituto de Tecnolog\'\i{}as en Detecci\'on y Astropart\'\i{}culas (CNEA, CONICET, UNSAM), Buenos Aires, Argentina}
\affiliation{International Center of Advanced Studies and Instituto de Ciencias F\'\i{}sicas, ECyT-UNSAM and CONICET, Campus Miguelete -- San Mart\'\i{}n, Buenos Aires, Argentina}
\affiliation{Laboratorio Atm\'osfera -- Departamento de Investigaciones en L\'aseres y sus Aplicaciones -- UNIDEF (CITEDEF-CONICET), Argentina}
\affiliation{Observatorio Pierre Auger, Malarg\"ue, Argentina}
\affiliation{Observatorio Pierre Auger and Comisi\'on Nacional de Energ\'\i{}a At\'omica, Malarg\"ue, Argentina}
\affiliation{Universidad Tecnol\'ogica Nacional -- Facultad Regional Buenos Aires, Buenos Aires, Argentina}
\affiliation{Adelaide University, Adelaide, S.A., Australia}
\affiliation{Universit\'e Libre de Bruxelles (ULB), Brussels, Belgium}
\affiliation{Vrije Universiteit Brussels, Brussels, Belgium}
\affiliation{Centro Brasileiro de Pesquisas Fisicas, Rio de Janeiro, RJ, Brazil}
\affiliation{Centro Federal de Educa\c{c}\~ao Tecnol\'ogica Celso Suckow da Fonseca, Petropolis, Brazil}
\affiliation{Universidade de S\~ao Paulo, Escola de Engenharia de Lorena, Lorena, SP, Brazil}
\affiliation{Universidade de S\~ao Paulo, Instituto de F\'\i{}sica de S\~ao Carlos, S\~ao Carlos, SP, Brazil}
\affiliation{Universidade de S\~ao Paulo, Instituto de F\'\i{}sica, S\~ao Paulo, SP, Brazil}
\affiliation{Universidade Estadual de Campinas (UNICAMP), IFGW, Campinas, SP, Brazil}
\affiliation{Universidade Estadual de Feira de Santana, Feira de Santana, Brazil}
\affiliation{Universidade Federal do ABC, Santo Andr\'e, SP, Brazil}
\affiliation{Universidade Federal do Paran\'a, Setor Palotina, Palotina, Brazil}
\affiliation{Universidade Federal do Rio de Janeiro, Instituto de F\'\i{}sica, Rio de Janeiro, RJ, Brazil}
\affiliation{Universidad de Medell\'\i{}n, Medell\'\i{}n, Colombia}
\affiliation{Universidad Industrial de Santander, Bucaramanga, Colombia}
\affiliation{Charles University, Faculty of Mathematics and Physics, Institute of Particle and Nuclear Physics, Prague, Czech Republic}
\affiliation{Institute of Physics of the Czech Academy of Sciences, Prague, Czech Republic}
\affiliation{Palacky University, Olomouc, Czech Republic}
\affiliation{CNRS/IN2P3, IJCLab, Universit\'e Paris-Saclay, Orsay, France}
\affiliation{Laboratoire de Physique Nucl\'eaire et de Hautes Energies (LPNHE), Sorbonne Universit\'e, Universit\'e de Paris, CNRS-IN2P3, Paris, France}
\affiliation{Universit\'e Paris-Saclay, CNRS/IN2P3, IJCLab, Orsay, France}
\affiliation{Bergische Universit\"at Wuppertal, Department of Physics, Wuppertal, Germany}
\affiliation{Karlsruhe Institute of Technology (KIT), Institute for Experimental Particle Physics, Karlsruhe, Germany}
\affiliation{Karlsruhe Institute of Technology (KIT), Institut f\"ur Prozessdatenverarbeitung und Elektronik, Karlsruhe, Germany}
\affiliation{Karlsruhe Institute of Technology (KIT), Institute for Astroparticle Physics, Karlsruhe, Germany}
\affiliation{RWTH Aachen University, III.\ Physikalisches Institut A, Aachen, Germany}
\affiliation{TU Dortmund University, Department of Physics, Dortmund, Germany}
\affiliation{Universit\"at Hamburg, II.\ Institut f\"ur Theoretische Physik, Hamburg, Germany}
\affiliation{Universit\"at Siegen, Department Physik -- Experimentelle Teilchenphysik, Siegen, Germany}
\affiliation{Gran Sasso Science Institute, L'Aquila, Italy}
\affiliation{INFN Laboratori Nazionali del Gran Sasso, Assergi (L'Aquila), Italy}
\affiliation{INFN, Sezione di Catania, Catania, Italy}
\affiliation{INFN, Sezione di Lecce, Lecce, Italy}
\affiliation{INFN, Sezione di Milano, Milano, Italy}
\affiliation{INFN, Sezione di Napoli, Napoli, Italy}
\affiliation{INFN, Sezione di Roma ``Tor Vergata'', Roma, Italy}
\affiliation{INFN, Sezione di Torino, Torino, Italy}
\affiliation{Osservatorio Astrofisico di Torino (INAF), Torino, Italy}
\affiliation{Politecnico di Milano, Dipartimento di Scienze e Tecnologie Aerospaziali , Milano, Italy}
\affiliation{Universit\`a del Salento, Dipartimento di Matematica e Fisica ``E.\ De Giorgi'', Lecce, Italy}
\affiliation{Universit\`a dell'Aquila, Dipartimento di Scienze Fisiche e Chimiche, L'Aquila, Italy}
\affiliation{Universit\`a di Catania, Dipartimento di Fisica e Astronomia ``Ettore Majorana``, Catania, Italy}
\affiliation{Universit\`a di Milano, Dipartimento di Fisica, Milano, Italy}
\affiliation{Universit\`a di Napoli ``Federico II'', Dipartimento di Fisica ``Ettore Pancini'', Napoli, Italy}
\affiliation{Universit\`a di Palermo, Dipartimento di Fisica e Chimica ''E.\ Segr\`e'', Palermo, Italy}
\affiliation{Universit\`a di Roma ``Tor Vergata'', Dipartimento di Fisica, Roma, Italy}
\affiliation{Universit\`a Torino, Dipartimento di Fisica, Torino, Italy}
\affiliation{Benem\'erita Universidad Aut\'onoma de Puebla, Puebla, M\'exico}
\affiliation{Unidad Profesional Interdisciplinaria en Ingenier\'\i{}a y Tecnolog\'\i{}as Avanzadas del Instituto Polit\'ecnico Nacional (UPIITA-IPN), M\'exico, D.F., M\'exico}
\affiliation{Universidad Aut\'onoma de Chiapas, Tuxtla Guti\'errez, Chiapas, M\'exico}
\affiliation{Universidad Michoacana de San Nicol\'as de Hidalgo, Morelia, Michoac\'an, M\'exico}
\affiliation{Universidad Nacional Aut\'onoma de M\'exico, M\'exico, D.F., M\'exico}
\affiliation{Institute of Nuclear Physics PAN, Krakow, Poland}
\affiliation{University of \L{}\'od\'z, Faculty of High-Energy Astrophysics,\L{}\'od\'z, Poland}
\affiliation{Laborat\'orio de Instrumenta\c{c}\~ao e F\'\i{}sica Experimental de Part\'\i{}culas -- LIP and Instituto Superior T\'ecnico -- IST, Universidade de Lisboa -- UL, Lisboa, Portugal}
\affiliation{``Horia Hulubei'' National Institute for Physics and Nuclear Engineering, Bucharest-Magurele, Romania}
\affiliation{Institute of Space Science, Bucharest-Magurele, Romania}
\affiliation{Center for Astrophysics and Cosmology (CAC), University of Nova Gorica, Nova Gorica, Slovenia}
\affiliation{Experimental Particle Physics Department, J.\ Stefan Institute, Ljubljana, Slovenia}
\affiliation{Instituto Galego de F\'\i{}sica de Altas Enerx\'\i{}as (IGFAE), Universidade de Santiago de Compostela, Santiago de Compostela, Spain}
\affiliation{IMAPP, Radboud University Nijmegen, Nijmegen, The Netherlands}
\affiliation{Nationaal Instituut voor Kernfysica en Hoge Energie Fysica (NIKHEF), Science Park, Amsterdam, The Netherlands}
\affiliation{Stichting Astronomisch Onderzoek in Nederland (ASTRON), Dwingeloo, The Netherlands}
\affiliation{Case Western Reserve University, Cleveland, OH, USA}
\affiliation{Colorado School of Mines, Golden, CO, USA}
\affiliation{Department of Physics and Astronomy, Lehman College, City University of New York, Bronx, NY, USA}
\affiliation{Michigan Technological University, Houghton, MI, USA}
\affiliation{New York University, New York, NY, USA}
\affiliation{University of Chicago, Enrico Fermi Institute, Chicago, IL, USA}
\affiliation{University of Delaware, Department of Physics and Astronomy, Bartol Research Institute, Newark, DE, USA}

\author{A.~\surname{Abdul Halim}}
\affiliation{Adelaide University, Adelaide, S.A., Australia}

\author{P.~\surname{Abreu}}
\affiliation{Laborat\'orio de Instrumenta\c{c}\~ao e F\'\i{}sica Experimental de Part\'\i{}culas -- LIP and Instituto Superior T\'ecnico -- IST, Universidade de Lisboa -- UL, Lisboa, Portugal}

\author{M.~\surname{Aglietta}}
\affiliation{Osservatorio Astrofisico di Torino (INAF), Torino, Italy}
\affiliation{INFN, Sezione di Torino, Torino, Italy}

\author{M.~\surname{Ahmed}}
\affiliation{CNRS/IN2P3, IJCLab, Universit\'e Paris-Saclay, Orsay, France}

\author{I.~\surname{Allekotte}}
\affiliation{Centro At\'omico Bariloche and Instituto Balseiro (CNEA-UNCuyo-CONICET), San Carlos de Bariloche, Argentina}

\author{K.~\surname{Almeida Cheminant}}
\affiliation{Nationaal Instituut voor Kernfysica en Hoge Energie Fysica (NIKHEF), Science Park, Amsterdam, The Netherlands}
\affiliation{IMAPP, Radboud University Nijmegen, Nijmegen, The Netherlands}

\author{R.~\surname{Aloisio}}
\affiliation{Gran Sasso Science Institute, L'Aquila, Italy}
\affiliation{INFN Laboratori Nazionali del Gran Sasso, Assergi (L'Aquila), Italy}

\author{J.~\surname{Alvarez-Mu\~niz}}
\affiliation{Instituto Galego de F\'\i{}sica de Altas Enerx\'\i{}as (IGFAE), Universidade de Santiago de Compostela, Santiago de Compostela, Spain}

\author{A.~\surname{Ambrosone}}
\affiliation{Gran Sasso Science Institute, L'Aquila, Italy}
\affiliation{INFN Laboratori Nazionali del Gran Sasso, Assergi (L'Aquila), Italy}

\author{J.~\surname{Ammerman Yebra}}
\affiliation{Instituto Galego de F\'\i{}sica de Altas Enerx\'\i{}as (IGFAE), Universidade de Santiago de Compostela, Santiago de Compostela, Spain}

% ...

\collaboration{The Pierre Auger Collaboration}
\email{spokespersons@auger.org}
\homepage{http://www.auger.org}
\noaffiliation

%% file: nuNtransitions.bbl
\begin{thebibliography}{59}%
\makeatletter
\providecommand \@ifxundefined [1]{%
 \@ifx{#1\undefined}
}%
\providecommand \@ifnum [1]{%
 \ifnum #1\expandafter \@firstoftwo
 \else \expandafter \@secondoftwo
 \fi
}%
\providecommand \@ifx [1]{%
 \ifx #1\expandafter \@firstoftwo
 \else \expandafter \@secondoftwo
 \fi
}%
\providecommand \natexlab [1]{#1}%
\providecommand \enquote  [1]{``#1''}%
\providecommand \bibnamefont  [1]{#1}%
\providecommand \bibfnamefont [1]{#1}%
\providecommand \citenamefont [1]{#1}%
\providecommand \href@noop [0]{\@secondoftwo}%
\providecommand \href [0]{\begingroup \@sanitize@url \@href}%
\providecommand \@href[1]{\@@startlink{#1}\@@href}%
\providecommand \@@href[1]{\endgroup#1\@@endlink}%
\providecommand \@sanitize@url [0]{\catcode `\\12\catcode `\$12\catcode
  `\&12\catcode `\#12\catcode `\^12\catcode `\_12\catcode `\%12\relax}%
\providecommand \@@startlink[1]{}%
\providecommand \@@endlink[0]{}%
\providecommand \url  [0]{\begingroup\@sanitize@url \@url }%
\providecommand \@url [1]{\endgroup\@href {#1}{\urlprefix }}%
\providecommand \urlprefix  [0]{URL }%
\providecommand \Eprint [0]{\href }%
\providecommand \doibase [0]{https://doi.org/}%
\providecommand \selectlanguage [0]{\@gobble}%
\providecommand \bibinfo  [0]{\@secondoftwo}%
\providecommand \bibfield  [0]{\@secondoftwo}%
\providecommand \translation [1]{[#1]}%
\providecommand \BibitemOpen [0]{}%
\providecommand \bibitemStop [0]{}%
\providecommand \bibitemNoStop [0]{.\EOS\space}%
\providecommand \EOS [0]{\spacefactor3000\relax}%
\providecommand \BibitemShut  [1]{\csname bibitem#1\endcsname}%
\let\auto@bib@innerbib\@empty
\bibitem [{\citenamefont {Abdul~Halim}\ \emph {et~al.}(2023)\citenamefont
  {Abdul~Halim} \emph {et~al.}}]{PierreAuger:2022atd}%
  \BibitemOpen
  \bibfield  {author} {\bibinfo {author} {\bibfnamefont {A.}~\bibnamefont
  {Abdul~Halim}} \emph {et~al.} (\bibinfo {collaboration} {Pierre Auger}),\
  }\href {https://doi.org/10.1088/1475-7516/2023/05/024} {\bibfield  {journal}
  {\bibinfo  {journal} {JCAP}\ }\textbf {\bibinfo {volume} {05}},\ \bibinfo
  {pages} {024 (2023)}},\ \Eprint {https://arxiv.org/abs/2211.02857}
  {arXiv:2211.02857 [astro-ph.HE]} \BibitemShut {NoStop}%
\bibitem [{\citenamefont {Aloisio}\ \emph {et~al.}(2014)\citenamefont
  {Aloisio}, \citenamefont {Berezinsky},\ and\ \citenamefont
  {Blasi}}]{Aloisio:2013hya}%
  \BibitemOpen
  \bibfield  {author} {\bibinfo {author} {\bibfnamefont {R.}~\bibnamefont
  {Aloisio}}, \bibinfo {author} {\bibfnamefont {V.}~\bibnamefont
  {Berezinsky}},\ and\ \bibinfo {author} {\bibfnamefont {P.}~\bibnamefont
  {Blasi}},\ }\href {https://doi.org/10.1088/1475-7516/2014/10/020} {\bibfield
  {journal} {\bibinfo  {journal} {JCAP}\ }\textbf {\bibinfo {volume} {10}},\
  \bibinfo {pages} {020 (2014)}},\ \Eprint {https://arxiv.org/abs/1312.7459}
  {arXiv:1312.7459 [astro-ph.HE]} \BibitemShut {NoStop}%
\bibitem [{\citenamefont {Biehl}\ \emph {et~al.}(2018)\citenamefont {Biehl},
  \citenamefont {Boncioli}, \citenamefont {Lunardini},\ and\ \citenamefont
  {Winter}}]{Biehl:2017hnb}%
  \BibitemOpen
  \bibfield  {author} {\bibinfo {author} {\bibfnamefont {D.}~\bibnamefont
  {Biehl}}, \bibinfo {author} {\bibfnamefont {D.}~\bibnamefont {Boncioli}},
  \bibinfo {author} {\bibfnamefont {C.}~\bibnamefont {Lunardini}},\ and\
  \bibinfo {author} {\bibfnamefont {W.}~\bibnamefont {Winter}},\ }\href
  {https://doi.org/10.1038/s41598-018-29022-4} {\bibfield  {journal} {\bibinfo
  {journal} {Sci. Rep.}\ }\textbf {\bibinfo {volume} {8}},\ \bibinfo {pages}
  {10828} (\bibinfo {year} {2018})},\ \Eprint
  {https://arxiv.org/abs/1711.03555} {arXiv:1711.03555 [astro-ph.HE]}
  \BibitemShut {NoStop}%
\bibitem [{\citenamefont {Alves~Batista}\ \emph {et~al.}(2019)\citenamefont
  {Alves~Batista}, \citenamefont {de~Almeida}, \citenamefont {Lago},\ and\
  \citenamefont {Kotera}}]{AlvesBatista:2018zui}%
  \BibitemOpen
  \bibfield  {author} {\bibinfo {author} {\bibfnamefont {R.}~\bibnamefont
  {Alves~Batista}}, \bibinfo {author} {\bibfnamefont {R.~M.}\ \bibnamefont
  {de~Almeida}}, \bibinfo {author} {\bibfnamefont {B.}~\bibnamefont {Lago}},\
  and\ \bibinfo {author} {\bibfnamefont {K.}~\bibnamefont {Kotera}},\ }\href
  {https://doi.org/10.1088/1475-7516/2019/01/002} {\bibfield  {journal}
  {\bibinfo  {journal} {JCAP}\ }\textbf {\bibinfo {volume} {01}},\ \bibinfo
  {pages} {002 (2019)}},\ \Eprint {https://arxiv.org/abs/1806.10879}
  {arXiv:1806.10879 [astro-ph.HE]} \BibitemShut {NoStop}%
\bibitem [{\citenamefont {Boncioli}\ \emph {et~al.}(2019)\citenamefont
  {Boncioli}, \citenamefont {Biehl},\ and\ \citenamefont
  {Winter}}]{Boncioli:2018lrv}%
  \BibitemOpen
  \bibfield  {author} {\bibinfo {author} {\bibfnamefont {D.}~\bibnamefont
  {Boncioli}}, \bibinfo {author} {\bibfnamefont {D.}~\bibnamefont {Biehl}},\
  and\ \bibinfo {author} {\bibfnamefont {W.}~\bibnamefont {Winter}},\ }\href
  {https://doi.org/10.3847/1538-4357/aafda7} {\bibfield  {journal} {\bibinfo
  {journal} {Astrophys. J.}\ }\textbf {\bibinfo {volume} {872}},\ \bibinfo
  {pages} {110} (\bibinfo {year} {2019})},\ \Eprint
  {https://arxiv.org/abs/1808.07481} {arXiv:1808.07481 [astro-ph.HE]}
  \BibitemShut {NoStop}%
\bibitem [{\citenamefont {Zhang}\ and\ \citenamefont
  {Murase}(2019)}]{Zhang:2018agl}%
  \BibitemOpen
  \bibfield  {author} {\bibinfo {author} {\bibfnamefont {B.~T.}\ \bibnamefont
  {Zhang}}\ and\ \bibinfo {author} {\bibfnamefont {K.}~\bibnamefont {Murase}},\
  }\href {https://doi.org/10.1103/PhysRevD.100.103004} {\bibfield  {journal}
  {\bibinfo  {journal} {Phys. Rev. D}\ }\textbf {\bibinfo {volume} {100}},\
  \bibinfo {pages} {103004} (\bibinfo {year} {2019})},\ \Eprint
  {https://arxiv.org/abs/1812.10289} {arXiv:1812.10289 [astro-ph.HE]}
  \BibitemShut {NoStop}%
\bibitem [{\citenamefont {Condorelli}\ \emph {et~al.}(2023)\citenamefont
  {Condorelli}, \citenamefont {Boncioli}, \citenamefont {Peretti},\ and\
  \citenamefont {Petrera}}]{Condorelli:2022vfa}%
  \BibitemOpen
  \bibfield  {author} {\bibinfo {author} {\bibfnamefont {A.}~\bibnamefont
  {Condorelli}}, \bibinfo {author} {\bibfnamefont {D.}~\bibnamefont
  {Boncioli}}, \bibinfo {author} {\bibfnamefont {E.}~\bibnamefont {Peretti}},\
  and\ \bibinfo {author} {\bibfnamefont {S.}~\bibnamefont {Petrera}},\ }\href
  {https://doi.org/10.1103/PhysRevD.107.083009} {\bibfield  {journal} {\bibinfo
   {journal} {Phys. Rev. D}\ }\textbf {\bibinfo {volume} {107}},\ \bibinfo
  {pages} {083009} (\bibinfo {year} {2023})},\ \Eprint
  {https://arxiv.org/abs/2209.08593} {arXiv:2209.08593 [astro-ph.HE]}
  \BibitemShut {NoStop}%
\bibitem [{\citenamefont {Muzio}\ and\ \citenamefont
  {Farrar}(2023)}]{Muzio:2022bak}%
  \BibitemOpen
  \bibfield  {author} {\bibinfo {author} {\bibfnamefont {M.~S.}\ \bibnamefont
  {Muzio}}\ and\ \bibinfo {author} {\bibfnamefont {G.~R.}\ \bibnamefont
  {Farrar}},\ }\href {https://doi.org/10.3847/2041-8213/acac93} {\bibfield
  {journal} {\bibinfo  {journal} {Astrophys. J. Lett.}\ }\textbf {\bibinfo
  {volume} {942}},\ \bibinfo {pages} {L39} (\bibinfo {year} {2023})},\ \Eprint
  {https://arxiv.org/abs/2209.08068} {arXiv:2209.08068 [astro-ph.HE]}
  \BibitemShut {NoStop}%
\bibitem [{\citenamefont {Rodrigues}\ \emph {et~al.}(2021)\citenamefont
  {Rodrigues}, \citenamefont {Heinze}, \citenamefont {Palladino}, \citenamefont
  {van Vliet},\ and\ \citenamefont {Winter}}]{Rodrigues:2020pli}%
  \BibitemOpen
  \bibfield  {author} {\bibinfo {author} {\bibfnamefont {X.}~\bibnamefont
  {Rodrigues}}, \bibinfo {author} {\bibfnamefont {J.}~\bibnamefont {Heinze}},
  \bibinfo {author} {\bibfnamefont {A.}~\bibnamefont {Palladino}}, \bibinfo
  {author} {\bibfnamefont {A.}~\bibnamefont {van Vliet}},\ and\ \bibinfo
  {author} {\bibfnamefont {W.}~\bibnamefont {Winter}},\ }\href
  {https://doi.org/10.1103/PhysRevLett.126.191101} {\bibfield  {journal}
  {\bibinfo  {journal} {Phys. Rev. Lett.}\ }\textbf {\bibinfo {volume} {126}},\
  \bibinfo {pages} {191101} (\bibinfo {year} {2021})},\ \Eprint
  {https://arxiv.org/abs/2003.08392} {arXiv:2003.08392 [astro-ph.HE]}
  \BibitemShut {NoStop}%
\bibitem [{\citenamefont {Muzio}\ \emph {et~al.}(2023)\citenamefont {Muzio},
  \citenamefont {Unger},\ and\ \citenamefont {Wissel}}]{Muzio:2023skc}%
  \BibitemOpen
  \bibfield  {author} {\bibinfo {author} {\bibfnamefont {M.~S.}\ \bibnamefont
  {Muzio}}, \bibinfo {author} {\bibfnamefont {M.}~\bibnamefont {Unger}},\ and\
  \bibinfo {author} {\bibfnamefont {S.}~\bibnamefont {Wissel}},\ }\href
  {https://doi.org/10.1103/PhysRevD.107.103030} {\bibfield  {journal} {\bibinfo
   {journal} {Phys. Rev. D}\ }\textbf {\bibinfo {volume} {107}},\ \bibinfo
  {pages} {103030} (\bibinfo {year} {2023})},\ \Eprint
  {https://arxiv.org/abs/2303.04170} {arXiv:2303.04170 [astro-ph.HE]}
  \BibitemShut {NoStop}%
\bibitem [{\citenamefont {Bérat}\ \emph {et~al.}(2024)\citenamefont {Bérat},
  \citenamefont {Condorelli}, \citenamefont {Deligny}, \citenamefont
  {Montanet},\ and\ \citenamefont {Torres}}]{Berat:2024rvf}%
  \BibitemOpen
  \bibfield  {author} {\bibinfo {author} {\bibfnamefont {C.}~\bibnamefont
  {Bérat}}, \bibinfo {author} {\bibfnamefont {A.}~\bibnamefont {Condorelli}},
  \bibinfo {author} {\bibfnamefont {O.}~\bibnamefont {Deligny}}, \bibinfo
  {author} {\bibfnamefont {F.}~\bibnamefont {Montanet}},\ and\ \bibinfo
  {author} {\bibfnamefont {Z.}~\bibnamefont {Torres}},\ }\href
  {https://doi.org/10.3847/1538-4357/ad372a} {\bibfield  {journal} {\bibinfo
  {journal} {Astrophys. J.}\ }\textbf {\bibinfo {volume} {966}},\ \bibinfo
  {pages} {186} (\bibinfo {year} {2024})},\ \Eprint
  {https://arxiv.org/abs/2402.04759} {arXiv:2402.04759 [astro-ph.HE]}
  \BibitemShut {NoStop}%
\bibitem [{\citenamefont {Botner}\ \emph {et~al.}(2005)\citenamefont {Botner}
  \emph {et~al.}}]{BOTNER2005367}%
  \BibitemOpen
  \bibfield  {author} {\bibinfo {author} {\bibfnamefont {O.}~\bibnamefont
  {Botner}} \emph {et~al.} (\bibinfo {collaboration} {IceCube}),\ }\href
  {https://doi.org/https://doi.org/10.1016/j.nuclphysbps.2005.01.132}
  {\bibfield  {journal} {\bibinfo  {journal} {Nuclear Physics B - Proceedings
  Supplements}\ }\textbf {\bibinfo {volume} {143}},\ \bibinfo {pages} {367}
  (\bibinfo {year} {2005})}\BibitemShut {NoStop}%
\bibitem [{\citenamefont {Aab}\ \emph {et~al.}(2015{\natexlab{a}})\citenamefont
  {Aab} \emph {et~al.}}]{PierreAuger:2015eyc}%
  \BibitemOpen
  \bibfield  {author} {\bibinfo {author} {\bibfnamefont {A.}~\bibnamefont
  {Aab}} \emph {et~al.} (\bibinfo {collaboration} {Pierre Auger}),\ }\href
  {https://doi.org/10.1016/j.nima.2015.06.058} {\bibfield  {journal} {\bibinfo
  {journal} {Nucl. Instrum. Meth. A}\ }\textbf {\bibinfo {volume} {798}},\
  \bibinfo {pages} {172} (\bibinfo {year} {2015}{\natexlab{a}})},\ \Eprint
  {https://arxiv.org/abs/1502.01323} {arXiv:1502.01323 [astro-ph.IM]}
  \BibitemShut {NoStop}%
\bibitem [{\citenamefont {Aiello}\ \emph {et~al.}(2025)\citenamefont {Aiello}
  \emph {et~al.}}]{KM3NeT:2025npi}%
  \BibitemOpen
  \bibfield  {author} {\bibinfo {author} {\bibfnamefont {S.}~\bibnamefont
  {Aiello}} \emph {et~al.} (\bibinfo {collaboration} {KM3NeT}),\ }\href
  {https://doi.org/10.1038/s41586-024-08543-1} {\bibfield  {journal} {\bibinfo
  {journal} {Nature}\ }\textbf {\bibinfo {volume} {638}},\ \bibinfo {pages}
  {376} (\bibinfo {year} {2025})},\ \bibinfo {note} {[Erratum: Nature 640, E3
  (2025)]}\BibitemShut {NoStop}%
\bibitem [{\citenamefont {Adriani}\ \emph {et~al.}(2025)\citenamefont {Adriani}
  \emph {et~al.}}]{KM3NeT:2025vut}%
  \BibitemOpen
  \bibfield  {author} {\bibinfo {author} {\bibfnamefont {O.}~\bibnamefont
  {Adriani}} \emph {et~al.} (\bibinfo {collaboration} {KM3NeT}),\ }\href
  {https://doi.org/10.3847/2041-8213/adcc29} {\bibfield  {journal} {\bibinfo
  {journal} {Astrophys. J. Lett.}\ }\textbf {\bibinfo {volume} {984}},\
  \bibinfo {pages} {L41} (\bibinfo {year} {2025})},\ \Eprint
  {https://arxiv.org/abs/2502.08508} {arXiv:2502.08508 [astro-ph.HE]}
  \BibitemShut {NoStop}%
\bibitem [{\citenamefont {Fujikawa}\ and\ \citenamefont
  {Shrock}(1980)}]{Fujikawa:1980yx}%
  \BibitemOpen
  \bibfield  {author} {\bibinfo {author} {\bibfnamefont {K.}~\bibnamefont
  {Fujikawa}}\ and\ \bibinfo {author} {\bibfnamefont {R.}~\bibnamefont
  {Shrock}},\ }\href {https://doi.org/10.1103/PhysRevLett.45.963} {\bibfield
  {journal} {\bibinfo  {journal} {Phys. Rev. Lett.}\ }\textbf {\bibinfo
  {volume} {45}},\ \bibinfo {pages} {963} (\bibinfo {year} {1980})}\BibitemShut
  {NoStop}%
\bibitem [{\citenamefont {Shrock}(1982)}]{Shrock:1982sc}%
  \BibitemOpen
  \bibfield  {author} {\bibinfo {author} {\bibfnamefont {R.~E.}\ \bibnamefont
  {Shrock}},\ }\href {https://doi.org/10.1016/0550-3213(82)90273-5} {\bibfield
  {journal} {\bibinfo  {journal} {Nucl. Phys. B}\ }\textbf {\bibinfo {volume}
  {206}},\ \bibinfo {pages} {359} (\bibinfo {year} {1982})}\BibitemShut
  {NoStop}%
\bibitem [{\citenamefont {Giunti}\ and\ \citenamefont
  {Studenikin}(2015)}]{Giunti:2014ixa}%
  \BibitemOpen
  \bibfield  {author} {\bibinfo {author} {\bibfnamefont {C.}~\bibnamefont
  {Giunti}}\ and\ \bibinfo {author} {\bibfnamefont {A.}~\bibnamefont
  {Studenikin}},\ }\href {https://doi.org/10.1103/RevModPhys.87.531} {\bibfield
   {journal} {\bibinfo  {journal} {Rev. Mod. Phys.}\ }\textbf {\bibinfo
  {volume} {87}},\ \bibinfo {pages} {531} (\bibinfo {year} {2015})},\ \Eprint
  {https://arxiv.org/abs/1403.6344} {arXiv:1403.6344 [hep-ph]} \BibitemShut
  {NoStop}%
\bibitem [{\citenamefont {Babu}\ \emph {et~al.}(2020)\citenamefont {Babu},
  \citenamefont {Jana},\ and\ \citenamefont {Lindner}}]{Babu:2020ivd}%
  \BibitemOpen
  \bibfield  {author} {\bibinfo {author} {\bibfnamefont {K.~S.}\ \bibnamefont
  {Babu}}, \bibinfo {author} {\bibfnamefont {S.}~\bibnamefont {Jana}},\ and\
  \bibinfo {author} {\bibfnamefont {M.}~\bibnamefont {Lindner}},\ }\href
  {https://doi.org/10.1007/JHEP10(2020)040} {\bibfield  {journal} {\bibinfo
  {journal} {JHEP}\ }\textbf {\bibinfo {volume} {10}},\ \bibinfo {pages} {040
  (2020)}},\ \Eprint {https://arxiv.org/abs/2007.04291} {arXiv:2007.04291
  [hep-ph]} \BibitemShut {NoStop}%
\bibitem [{\citenamefont {Kusenko}\ and\ \citenamefont
  {Weiler}(2002)}]{Kusenko:2001gj}%
  \BibitemOpen
  \bibfield  {author} {\bibinfo {author} {\bibfnamefont {A.}~\bibnamefont
  {Kusenko}}\ and\ \bibinfo {author} {\bibfnamefont {T.~J.}\ \bibnamefont
  {Weiler}},\ }\href {https://doi.org/10.1103/PhysRevLett.88.161101} {\bibfield
   {journal} {\bibinfo  {journal} {Phys. Rev. Lett.}\ }\textbf {\bibinfo
  {volume} {88}},\ \bibinfo {pages} {161101} (\bibinfo {year} {2002})},\
  \Eprint {https://arxiv.org/abs/hep-ph/0106071} {arXiv:hep-ph/0106071}
  \BibitemShut {NoStop}%
\bibitem [{\citenamefont {Anchordoqui}\ \emph {et~al.}(2002)\citenamefont
  {Anchordoqui}, \citenamefont {Feng}, \citenamefont {Goldberg},\ and\
  \citenamefont {Shapere}}]{Anchordoqui:2001cg}%
  \BibitemOpen
  \bibfield  {author} {\bibinfo {author} {\bibfnamefont {L.~A.}\ \bibnamefont
  {Anchordoqui}}, \bibinfo {author} {\bibfnamefont {J.~L.}\ \bibnamefont
  {Feng}}, \bibinfo {author} {\bibfnamefont {H.}~\bibnamefont {Goldberg}},\
  and\ \bibinfo {author} {\bibfnamefont {A.~D.}\ \bibnamefont {Shapere}},\
  }\href {https://doi.org/10.1103/PhysRevD.65.124027} {\bibfield  {journal}
  {\bibinfo  {journal} {Phys. Rev. D}\ }\textbf {\bibinfo {volume} {65}},\
  \bibinfo {pages} {124027} (\bibinfo {year} {2002})},\ \Eprint
  {https://arxiv.org/abs/hep-ph/0112247} {arXiv:hep-ph/0112247} \BibitemShut
  {NoStop}%
\bibitem [{\citenamefont {Anchordoqui}\ \emph {et~al.}(2010)\citenamefont
  {Anchordoqui}, \citenamefont {Goldberg}, \citenamefont {Gora}, \citenamefont
  {Paul}, \citenamefont {Roth}, \citenamefont {Sarkar},\ and\ \citenamefont
  {Winders}}]{Anchordoqui:2010hq}%
  \BibitemOpen
  \bibfield  {author} {\bibinfo {author} {\bibfnamefont {L.~A.}\ \bibnamefont
  {Anchordoqui}}, \bibinfo {author} {\bibfnamefont {H.}~\bibnamefont
  {Goldberg}}, \bibinfo {author} {\bibfnamefont {D.}~\bibnamefont {Gora}},
  \bibinfo {author} {\bibfnamefont {T.}~\bibnamefont {Paul}}, \bibinfo {author}
  {\bibfnamefont {M.}~\bibnamefont {Roth}}, \bibinfo {author} {\bibfnamefont
  {S.}~\bibnamefont {Sarkar}},\ and\ \bibinfo {author} {\bibfnamefont {L.~L.}\
  \bibnamefont {Winders}},\ }\href {https://doi.org/10.1103/PhysRevD.82.043001}
  {\bibfield  {journal} {\bibinfo  {journal} {Phys. Rev. D}\ }\textbf {\bibinfo
  {volume} {82}},\ \bibinfo {pages} {043001} (\bibinfo {year} {2010})},\
  \Eprint {https://arxiv.org/abs/1004.3190} {arXiv:1004.3190 [hep-ph]}
  \BibitemShut {NoStop}%
\bibitem [{\citenamefont {Capelle}\ \emph {et~al.}(1998)\citenamefont
  {Capelle}, \citenamefont {Cronin}, \citenamefont {Parente},\ and\
  \citenamefont {Zas}}]{Capelle:1998zz}%
  \BibitemOpen
  \bibfield  {author} {\bibinfo {author} {\bibfnamefont {K.~S.}\ \bibnamefont
  {Capelle}}, \bibinfo {author} {\bibfnamefont {J.~W.}\ \bibnamefont {Cronin}},
  \bibinfo {author} {\bibfnamefont {G.}~\bibnamefont {Parente}},\ and\ \bibinfo
  {author} {\bibfnamefont {E.}~\bibnamefont {Zas}},\ }\href
  {https://doi.org/10.1016/S0927-6505(97)00059-5} {\bibfield  {journal}
  {\bibinfo  {journal} {Astropart. Phys.}\ }\textbf {\bibinfo {volume} {8}},\
  \bibinfo {pages} {321} (\bibinfo {year} {1998})},\ \Eprint
  {https://arxiv.org/abs/astro-ph/9801313} {arXiv:astro-ph/9801313}
  \BibitemShut {NoStop}%
\bibitem [{\citenamefont {Bertou}\ \emph {et~al.}(2002)\citenamefont {Bertou},
  \citenamefont {Billoir}, \citenamefont {Deligny}, \citenamefont {Lachaud},\
  and\ \citenamefont {Letessier-Selvon}}]{Bertou:2001vm}%
  \BibitemOpen
  \bibfield  {author} {\bibinfo {author} {\bibfnamefont {X.}~\bibnamefont
  {Bertou}}, \bibinfo {author} {\bibfnamefont {P.}~\bibnamefont {Billoir}},
  \bibinfo {author} {\bibfnamefont {O.}~\bibnamefont {Deligny}}, \bibinfo
  {author} {\bibfnamefont {C.}~\bibnamefont {Lachaud}},\ and\ \bibinfo {author}
  {\bibfnamefont {A.}~\bibnamefont {Letessier-Selvon}},\ }\href
  {https://doi.org/10.1016/S0927-6505(01)00147-5} {\bibfield  {journal}
  {\bibinfo  {journal} {Astropart. Phys.}\ }\textbf {\bibinfo {volume} {17}},\
  \bibinfo {pages} {183} (\bibinfo {year} {2002})},\ \Eprint
  {https://arxiv.org/abs/astro-ph/0104452} {arXiv:astro-ph/0104452}
  \BibitemShut {NoStop}%
\bibitem [{\citenamefont {Domokos}\ and\ \citenamefont
  {Kovesi-Domokos}(1997)}]{Domokos:1996cn}%
  \BibitemOpen
  \bibfield  {author} {\bibinfo {author} {\bibfnamefont {G.}~\bibnamefont
  {Domokos}}\ and\ \bibinfo {author} {\bibfnamefont {S.}~\bibnamefont
  {Kovesi-Domokos}},\ }\href {https://doi.org/10.1103/PhysRevD.55.R2526}
  {\bibfield  {journal} {\bibinfo  {journal} {Phys. Rev. D}\ }\textbf {\bibinfo
  {volume} {55}},\ \bibinfo {pages} {2526} (\bibinfo {year} {1997})},\ \Eprint
  {https://arxiv.org/abs/hep-ph/9603242} {arXiv:hep-ph/9603242} \BibitemShut
  {NoStop}%
\bibitem [{\citenamefont {Buckley}\ \emph {et~al.}(2015)\citenamefont
  {Buckley}, \citenamefont {Ferrando}, \citenamefont {Lloyd}, \citenamefont
  {Nordstr{\"o}m}, \citenamefont {Page}, \citenamefont {R{\"u}fenacht},
  \citenamefont {Sch{\"o}nherr},\ and\ \citenamefont {Watt}}]{Buckley:2014ana}%
  \BibitemOpen
  \bibfield  {author} {\bibinfo {author} {\bibfnamefont {A.}~\bibnamefont
  {Buckley}}, \bibinfo {author} {\bibfnamefont {J.}~\bibnamefont {Ferrando}},
  \bibinfo {author} {\bibfnamefont {S.}~\bibnamefont {Lloyd}}, \bibinfo
  {author} {\bibfnamefont {K.}~\bibnamefont {Nordstr{\"o}m}}, \bibinfo {author}
  {\bibfnamefont {B.}~\bibnamefont {Page}}, \bibinfo {author} {\bibfnamefont
  {M.}~\bibnamefont {R{\"u}fenacht}}, \bibinfo {author} {\bibfnamefont
  {M.}~\bibnamefont {Sch{\"o}nherr}},\ and\ \bibinfo {author} {\bibfnamefont
  {G.}~\bibnamefont {Watt}},\ }\href
  {https://doi.org/10.1140/epjc/s10052-015-3318-8} {\bibfield  {journal}
  {\bibinfo  {journal} {Eur. Phys. J. C}\ }\textbf {\bibinfo {volume} {75}},\
  \bibinfo {pages} {132} (\bibinfo {year} {2015})},\ \Eprint
  {https://arxiv.org/abs/1412.7420} {arXiv:1412.7420 [hep-ph]} \BibitemShut
  {NoStop}%
\bibitem [{\citenamefont {Gandhi}\ \emph {et~al.}(1996)\citenamefont {Gandhi},
  \citenamefont {Quigg}, \citenamefont {Reno},\ and\ \citenamefont
  {Sarcevic}}]{Gandhi:1995tf}%
  \BibitemOpen
  \bibfield  {author} {\bibinfo {author} {\bibfnamefont {R.}~\bibnamefont
  {Gandhi}}, \bibinfo {author} {\bibfnamefont {C.}~\bibnamefont {Quigg}},
  \bibinfo {author} {\bibfnamefont {M.~H.}\ \bibnamefont {Reno}},\ and\
  \bibinfo {author} {\bibfnamefont {I.}~\bibnamefont {Sarcevic}},\ }\href
  {https://doi.org/10.1016/0927-6505(96)00008-4} {\bibfield  {journal}
  {\bibinfo  {journal} {Astropart. Phys.}\ }\textbf {\bibinfo {volume} {5}},\
  \bibinfo {pages} {81} (\bibinfo {year} {1996})},\ \Eprint
  {https://arxiv.org/abs/hep-ph/9512364} {arXiv:hep-ph/9512364} \BibitemShut
  {NoStop}%
\bibitem [{\citenamefont {Armesto}\ \emph {et~al.}(2008)\citenamefont
  {Armesto}, \citenamefont {Merino}, \citenamefont {Parente},\ and\
  \citenamefont {Zas}}]{Armesto:2007tg}%
  \BibitemOpen
  \bibfield  {author} {\bibinfo {author} {\bibfnamefont {N.}~\bibnamefont
  {Armesto}}, \bibinfo {author} {\bibfnamefont {C.}~\bibnamefont {Merino}},
  \bibinfo {author} {\bibfnamefont {G.}~\bibnamefont {Parente}},\ and\ \bibinfo
  {author} {\bibfnamefont {E.}~\bibnamefont {Zas}},\ }\href
  {https://doi.org/10.1103/PhysRevD.77.013001} {\bibfield  {journal} {\bibinfo
  {journal} {Phys. Rev. D}\ }\textbf {\bibinfo {volume} {77}},\ \bibinfo
  {pages} {013001} (\bibinfo {year} {2008})},\ \Eprint
  {https://arxiv.org/abs/0709.4461} {arXiv:0709.4461 [hep-ph]} \BibitemShut
  {NoStop}%
\bibitem [{\citenamefont {Connolly}\ \emph {et~al.}(2011)\citenamefont
  {Connolly}, \citenamefont {Thorne},\ and\ \citenamefont
  {Waters}}]{Connolly:2011vc}%
  \BibitemOpen
  \bibfield  {author} {\bibinfo {author} {\bibfnamefont {A.}~\bibnamefont
  {Connolly}}, \bibinfo {author} {\bibfnamefont {R.~S.}\ \bibnamefont
  {Thorne}},\ and\ \bibinfo {author} {\bibfnamefont {D.}~\bibnamefont
  {Waters}},\ }\href {https://doi.org/10.1103/PhysRevD.83.113009} {\bibfield
  {journal} {\bibinfo  {journal} {Phys. Rev. D}\ }\textbf {\bibinfo {volume}
  {83}},\ \bibinfo {pages} {113009} (\bibinfo {year} {2011})},\ \Eprint
  {https://arxiv.org/abs/1102.0691} {arXiv:1102.0691 [hep-ph]} \BibitemShut
  {NoStop}%
\bibitem [{\citenamefont {Abbasi}\ \emph {et~al.}(2021)\citenamefont {Abbasi}
  \emph {et~al.}}]{IceCube:2020rnc}%
  \BibitemOpen
  \bibfield  {author} {\bibinfo {author} {\bibfnamefont {R.}~\bibnamefont
  {Abbasi}} \emph {et~al.} (\bibinfo {collaboration} {IceCube}),\ }\href
  {https://doi.org/10.1103/PhysRevD.104.022001} {\bibfield  {journal} {\bibinfo
   {journal} {Phys. Rev. D}\ }\textbf {\bibinfo {volume} {104}},\ \bibinfo
  {pages} {022001} (\bibinfo {year} {2021})},\ \Eprint
  {https://arxiv.org/abs/2011.03560} {arXiv:2011.03560 [hep-ex]} \BibitemShut
  {NoStop}%
\bibitem [{\citenamefont {Aab}\ \emph {et~al.}(2019)\citenamefont {Aab} \emph
  {et~al.}}]{PierreAuger:2019ens}%
  \BibitemOpen
  \bibfield  {author} {\bibinfo {author} {\bibfnamefont {A.}~\bibnamefont
  {Aab}} \emph {et~al.} (\bibinfo {collaboration} {Pierre Auger}),\ }\href
  {https://doi.org/10.1088/1475-7516/2019/10/022} {\bibfield  {journal}
  {\bibinfo  {journal} {JCAP}\ }\textbf {\bibinfo {volume} {10}},\ \bibinfo
  {pages} {022 (2019)}},\ \Eprint {https://arxiv.org/abs/1906.07422}
  {arXiv:1906.07422 [astro-ph.HE]} \BibitemShut {NoStop}%
\bibitem [{\citenamefont {Abreu}\ \emph {et~al.}(2011)\citenamefont {Abreu}
  \emph {et~al.}}]{PierreAuger:2011cpc}%
  \BibitemOpen
  \bibfield  {author} {\bibinfo {author} {\bibfnamefont {P.}~\bibnamefont
  {Abreu}} \emph {et~al.} (\bibinfo {collaboration} {Pierre Auger}),\ }\href
  {https://doi.org/10.1103/PhysRevD.84.122005} {\bibfield  {journal} {\bibinfo
  {journal} {Phys. Rev. D}\ }\textbf {\bibinfo {volume} {84}},\ \bibinfo
  {pages} {122005} (\bibinfo {year} {2011})},\ \bibinfo {note} {[Erratum:
  Phys.Rev.D 84, 029902 (2011)]},\ \Eprint {https://arxiv.org/abs/1202.1493}
  {arXiv:1202.1493 [astro-ph.HE]} \BibitemShut {NoStop}%
\bibitem [{\citenamefont {Aab}\ \emph {et~al.}(2015{\natexlab{b}})\citenamefont
  {Aab} \emph {et~al.}}]{PierreAuger:2015ihf}%
  \BibitemOpen
  \bibfield  {author} {\bibinfo {author} {\bibfnamefont {A.}~\bibnamefont
  {Aab}} \emph {et~al.} (\bibinfo {collaboration} {Pierre Auger}),\ }\href
  {https://doi.org/10.1103/PhysRevD.91.092008} {\bibfield  {journal} {\bibinfo
  {journal} {Phys. Rev. D}\ }\textbf {\bibinfo {volume} {91}},\ \bibinfo
  {pages} {092008} (\bibinfo {year} {2015}{\natexlab{b}})},\ \Eprint
  {https://arxiv.org/abs/1504.05397} {arXiv:1504.05397 [astro-ph.HE]}
  \BibitemShut {NoStop}%
\bibitem [{\citenamefont {Zhang}\ and\ \citenamefont
  {Liu}(2023)}]{Zhang:2023nxy}%
  \BibitemOpen
  \bibfield  {author} {\bibinfo {author} {\bibfnamefont {Y.}~\bibnamefont
  {Zhang}}\ and\ \bibinfo {author} {\bibfnamefont {W.}~\bibnamefont {Liu}},\
  }\href {https://doi.org/10.1103/PhysRevD.107.095031} {\bibfield  {journal}
  {\bibinfo  {journal} {Phys. Rev. D}\ }\textbf {\bibinfo {volume} {107}},\
  \bibinfo {pages} {095031} (\bibinfo {year} {2023})},\ \Eprint
  {https://arxiv.org/abs/2301.06050} {arXiv:2301.06050 [hep-ph]} \BibitemShut
  {NoStop}%
\bibitem [{\citenamefont {Ovchynnikov}\ and\ \citenamefont
  {Zhu}(2023)}]{Ovchynnikov:2023wgg}%
  \BibitemOpen
  \bibfield  {author} {\bibinfo {author} {\bibfnamefont {M.}~\bibnamefont
  {Ovchynnikov}}\ and\ \bibinfo {author} {\bibfnamefont {J.-Y.}\ \bibnamefont
  {Zhu}},\ }\href {https://doi.org/10.1007/JHEP07(2023)039} {\bibfield
  {journal} {\bibinfo  {journal} {JHEP}\ }\textbf {\bibinfo {volume} {07}},\
  \bibinfo {pages} {039}},\ \Eprint {https://arxiv.org/abs/2301.08592}
  {arXiv:2301.08592 [hep-ph]} \BibitemShut {NoStop}%
\bibitem [{\citenamefont {Brdar}\ \emph {et~al.}(2025)\citenamefont {Brdar},
  \citenamefont {Li}, \citenamefont {Mir},\ and\ \citenamefont
  {Wang}}]{Brdar:2025iua}%
  \BibitemOpen
  \bibfield  {author} {\bibinfo {author} {\bibfnamefont {V.}~\bibnamefont
  {Brdar}}, \bibinfo {author} {\bibfnamefont {Y.-Y.}\ \bibnamefont {Li}},
  \bibinfo {author} {\bibfnamefont {S.~R.}\ \bibnamefont {Mir}},\ and\ \bibinfo
  {author} {\bibfnamefont {Y.-L.}\ \bibnamefont {Wang}},\ }\href
  {https://doi.org/10.1007/JHEP10(2025)230} {\bibfield  {journal} {\bibinfo
  {journal} {JHEP}\ }\textbf {\bibinfo {volume} {10}},\ \bibinfo {pages} {230
  (2025)}},\ \Eprint {https://arxiv.org/abs/2502.07024} {arXiv:2502.07024
  [hep-ph]} \BibitemShut {NoStop}%
\bibitem [{\citenamefont {Cornwall}\ \emph {et~al.}(1974)\citenamefont
  {Cornwall}, \citenamefont {Levin},\ and\ \citenamefont
  {Tiktopoulos}}]{Cornwall:1974km}%
  \BibitemOpen
  \bibfield  {author} {\bibinfo {author} {\bibfnamefont {J.~M.}\ \bibnamefont
  {Cornwall}}, \bibinfo {author} {\bibfnamefont {D.~N.}\ \bibnamefont
  {Levin}},\ and\ \bibinfo {author} {\bibfnamefont {G.}~\bibnamefont
  {Tiktopoulos}},\ }\href {https://doi.org/10.1103/PhysRevD.10.1145} {\bibfield
   {journal} {\bibinfo  {journal} {Phys. Rev. D}\ }\textbf {\bibinfo {volume}
  {10}},\ \bibinfo {pages} {1145} (\bibinfo {year} {1974})},\ \bibinfo {note}
  {[Erratum: Phys.Rev.D 11, 972 (1975)]}\BibitemShut {NoStop}%
\bibitem [{\citenamefont {Anchordoqui}\ and\ \citenamefont
  {Vafa}(2026)}]{Anchordoqui:2026ret}%
  \BibitemOpen
  \bibfield  {author} {\bibinfo {author} {\bibfnamefont {L.~A.}\ \bibnamefont
  {Anchordoqui}}\ and\ \bibinfo {author} {\bibfnamefont {C.}~\bibnamefont
  {Vafa}},\ }\href@noop {} {\  (\bibinfo {year} {2026})},\ \Eprint
  {https://arxiv.org/abs/2610.15209} {arXiv:2610.15209 [hep-ph]} \BibitemShut
  {NoStop}%
\bibitem [{\citenamefont {Blennow}\ \emph {et~al.}(2023)\citenamefont
  {Blennow}, \citenamefont {Fern{\'a}ndez-Mart{\'\i}nez}, \citenamefont
  {Hern{\'a}ndez-Garc{\'\i}a}, \citenamefont {L{\'o}pez-Pav{\'o}n},
  \citenamefont {Marcano},\ and\ \citenamefont
  {Naredo-Tuero}}]{Blennow:2023mqx}%
  \BibitemOpen
  \bibfield  {author} {\bibinfo {author} {\bibfnamefont {M.}~\bibnamefont
  {Blennow}}, \bibinfo {author} {\bibfnamefont {E.}~\bibnamefont
  {Fern{\'a}ndez-Mart{\'\i}nez}}, \bibinfo {author} {\bibfnamefont
  {J.}~\bibnamefont {Hern{\'a}ndez-Garc{\'\i}a}}, \bibinfo {author}
  {\bibfnamefont {J.}~\bibnamefont {L{\'o}pez-Pav{\'o}n}}, \bibinfo {author}
  {\bibfnamefont {X.}~\bibnamefont {Marcano}},\ and\ \bibinfo {author}
  {\bibfnamefont {D.}~\bibnamefont {Naredo-Tuero}},\ }\href
  {https://doi.org/10.1007/JHEP08(2023)030} {\bibfield  {journal} {\bibinfo
  {journal} {JHEP}\ }\textbf {\bibinfo {volume} {08}},\ \bibinfo {pages}
  {030}},\ \Eprint {https://arxiv.org/abs/2306.01040} {arXiv:2306.01040
  [hep-ph]} \BibitemShut {NoStop}%
\bibitem [{\citenamefont {Bauer}\ \emph {et~al.}(2021)\citenamefont {Bauer},
  \citenamefont {Rodd},\ and\ \citenamefont {Webber}}]{Bauer:2020jay}%
  \BibitemOpen
  \bibfield  {author} {\bibinfo {author} {\bibfnamefont {C.~W.}\ \bibnamefont
  {Bauer}}, \bibinfo {author} {\bibfnamefont {N.~L.}\ \bibnamefont {Rodd}},\
  and\ \bibinfo {author} {\bibfnamefont {B.~R.}\ \bibnamefont {Webber}},\
  }\href {https://doi.org/10.1007/JHEP06(2021)121} {\bibfield  {journal}
  {\bibinfo  {journal} {JHEP}\ }\textbf {\bibinfo {volume} {06}},\ \bibinfo
  {pages} {121 (2021)}},\ \Eprint {https://arxiv.org/abs/2007.15001}
  {arXiv:2007.15001 [hep-ph]} \BibitemShut {NoStop}%
\bibitem [{\citenamefont {Abraham}\ \emph {et~al.}(2009)\citenamefont {Abraham}
  \emph {et~al.}}]{PierreAuger:2009dvq}%
  \BibitemOpen
  \bibfield  {author} {\bibinfo {author} {\bibfnamefont {J.}~\bibnamefont
  {Abraham}} \emph {et~al.} (\bibinfo {collaboration} {Pierre Auger}),\ }\href
  {https://doi.org/10.1103/PhysRevD.79.102001} {\bibfield  {journal} {\bibinfo
  {journal} {Phys. Rev. D}\ }\textbf {\bibinfo {volume} {79}},\ \bibinfo
  {pages} {102001} (\bibinfo {year} {2009})},\ \Eprint
  {https://arxiv.org/abs/0903.3385} {arXiv:0903.3385 [astro-ph.HE]}
  \BibitemShut {NoStop}%
\bibitem [{\citenamefont {Corcella}\ \emph {et~al.}(2001)\citenamefont
  {Corcella}, \citenamefont {Knowles}, \citenamefont {Marchesini},
  \citenamefont {Moretti}, \citenamefont {Odagiri}, \citenamefont {Richardson},
  \citenamefont {Seymour},\ and\ \citenamefont {Webber}}]{Corcella:2000bw}%
  \BibitemOpen
  \bibfield  {author} {\bibinfo {author} {\bibfnamefont {G.}~\bibnamefont
  {Corcella}}, \bibinfo {author} {\bibfnamefont {I.}~\bibnamefont {Knowles}},
  \bibinfo {author} {\bibfnamefont {G.}~\bibnamefont {Marchesini}}, \bibinfo
  {author} {\bibfnamefont {S.}~\bibnamefont {Moretti}}, \bibinfo {author}
  {\bibfnamefont {K.}~\bibnamefont {Odagiri}}, \bibinfo {author} {\bibfnamefont
  {P.}~\bibnamefont {Richardson}}, \bibinfo {author} {\bibfnamefont
  {M.}~\bibnamefont {Seymour}},\ and\ \bibinfo {author} {\bibfnamefont
  {B.}~\bibnamefont {Webber}},\ }\href
  {https://doi.org/10.1088/1126-6708/2001/01/010} {\bibfield  {journal}
  {\bibinfo  {journal} {JHEP}\ }\textbf {\bibinfo {volume} {01}},\ \bibinfo
  {pages} {010}},\ \Eprint {https://arxiv.org/abs/hep-ph/0011363}
  {hep-ph/0011363} \BibitemShut {NoStop}%
\bibitem [{\citenamefont {Sjostrand}\ \emph {et~al.}(2006)\citenamefont
  {Sjostrand}, \citenamefont {Mrenna},\ and\ \citenamefont
  {Skands}}]{Sjostrand:2006za}%
  \BibitemOpen
  \bibfield  {author} {\bibinfo {author} {\bibfnamefont {T.}~\bibnamefont
  {Sjostrand}}, \bibinfo {author} {\bibfnamefont {S.}~\bibnamefont {Mrenna}},\
  and\ \bibinfo {author} {\bibfnamefont {P.~Z.}\ \bibnamefont {Skands}},\
  }\href {https://doi.org/10.1088/1126-6708/2006/05/026} {\bibfield  {journal}
  {\bibinfo  {journal} {JHEP}\ }\textbf {\bibinfo {volume} {05}},\ \bibinfo
  {pages} {026}},\ \Eprint {https://arxiv.org/abs/hep-ph/0603175}
  {hep-ph/0603175} \BibitemShut {NoStop}%
\bibitem [{\citenamefont {Sciutto}(1999)}]{Sciutto:1999xz}%
  \BibitemOpen
  \bibfield  {author} {\bibinfo {author} {\bibfnamefont {S.~J.}\ \bibnamefont
  {Sciutto}},\ }\href@noop {} {\  (\bibinfo {year} {1999})},\ \Eprint
  {https://arxiv.org/abs/astro-ph/9911331} {arXiv:astro-ph/9911331}
  \BibitemShut {NoStop}%
\bibitem [{\citenamefont {Heck}\ \emph {et~al.}(1998)\citenamefont {Heck},
  \citenamefont {Knapp}, \citenamefont {Capdevielle}, \citenamefont {Schatz},\
  and\ \citenamefont {Thouw}}]{Heck:1998vt}%
  \BibitemOpen
  \bibfield  {author} {\bibinfo {author} {\bibfnamefont {D.}~\bibnamefont
  {Heck}}, \bibinfo {author} {\bibfnamefont {J.}~\bibnamefont {Knapp}},
  \bibinfo {author} {\bibfnamefont {J.}~\bibnamefont {Capdevielle}}, \bibinfo
  {author} {\bibfnamefont {G.}~\bibnamefont {Schatz}},\ and\ \bibinfo {author}
  {\bibfnamefont {T.}~\bibnamefont {Thouw}},\ }\href@noop {} {\bibinfo {title}
  {{CORSIKA: A Monte Carlo code to simulate extensive air showers}}} (\bibinfo
  {year} {1998}),\ \bibinfo {note} {forschungszentrum Karlsruhe Report FZKA
  6019}\BibitemShut {NoStop}%
\bibitem [{\citenamefont {Fletcher}\ \emph {et~al.}(1994)\citenamefont
  {Fletcher}, \citenamefont {Gaisser}, \citenamefont {Lipari},\ and\
  \citenamefont {Stanev}}]{Fletcher:1994bd}%
  \BibitemOpen
  \bibfield  {author} {\bibinfo {author} {\bibfnamefont {R.}~\bibnamefont
  {Fletcher}}, \bibinfo {author} {\bibfnamefont {T.}~\bibnamefont {Gaisser}},
  \bibinfo {author} {\bibfnamefont {P.}~\bibnamefont {Lipari}},\ and\ \bibinfo
  {author} {\bibfnamefont {T.}~\bibnamefont {Stanev}},\ }\href
  {https://doi.org/10.1103/PhysRevD.50.5710} {\bibfield  {journal} {\bibinfo
  {journal} {Phys. Rev. D}\ }\textbf {\bibinfo {volume} {50}},\ \bibinfo
  {pages} {5710} (\bibinfo {year} {1994})}\BibitemShut {NoStop}%
\bibitem [{\citenamefont {Kalmykov}\ \emph {et~al.}(1997)\citenamefont
  {Kalmykov}, \citenamefont {Ostapchenko},\ and\ \citenamefont
  {Pavlov}}]{Kalmykov:1997te}%
  \BibitemOpen
  \bibfield  {author} {\bibinfo {author} {\bibfnamefont {N.}~\bibnamefont
  {Kalmykov}}, \bibinfo {author} {\bibfnamefont {S.}~\bibnamefont
  {Ostapchenko}},\ and\ \bibinfo {author} {\bibfnamefont {A.}~\bibnamefont
  {Pavlov}},\ }\href {https://doi.org/10.1016/S0920-5632(96)00846-8} {\bibfield
   {journal} {\bibinfo  {journal} {Nucl. Phys. B Proc. Suppl.}\ }\textbf
  {\bibinfo {volume} {52}},\ \bibinfo {pages} {17} (\bibinfo {year}
  {1997})}\BibitemShut {NoStop}%
\bibitem [{\citenamefont {Hillas}(1997)}]{Hillas:1997tf}%
  \BibitemOpen
  \bibfield  {author} {\bibinfo {author} {\bibfnamefont {A.~M.}\ \bibnamefont
  {Hillas}},\ }\href {https://doi.org/10.1016/S0920-5632(96)00847-X} {\bibfield
   {journal} {\bibinfo  {journal} {Nucl. Phys. B Proc. Suppl.}\ }\textbf
  {\bibinfo {volume} {52}},\ \bibinfo {pages} {29} (\bibinfo {year}
  {1997})}\BibitemShut {NoStop}%
\bibitem [{\citenamefont {Cooper-Sarkar}\ and\ \citenamefont
  {Sarkar}(2008)}]{Cooper-Sarkar:2007zsa}%
  \BibitemOpen
  \bibfield  {author} {\bibinfo {author} {\bibfnamefont {A.}~\bibnamefont
  {Cooper-Sarkar}}\ and\ \bibinfo {author} {\bibfnamefont {S.}~\bibnamefont
  {Sarkar}},\ }\href {https://doi.org/10.1088/1126-6708/2008/01/075} {\bibfield
   {journal} {\bibinfo  {journal} {JHEP}\ }\textbf {\bibinfo {volume} {01}},\
  \bibinfo {pages} {075 (2008)}},\ \Eprint {https://arxiv.org/abs/0710.5303}
  {arXiv:0710.5303 [hep-ph]} \BibitemShut {NoStop}%
\bibitem [{\citenamefont {Gupta~(ed.)}(2021)}]{EarthGeophysics}%
  \BibitemOpen
  \bibfield  {author} {\bibinfo {author} {\bibfnamefont {H.~K.}\ \bibnamefont
  {Gupta~(ed.)}},\ }\href@noop {} {\emph {\bibinfo {title} {{Encyclopedia of
  solid Earth geophysics, 2nd edition}}}}\ (\bibinfo  {publisher} {Springer},\
  \bibinfo {year} {2021})\BibitemShut {NoStop}%
\bibitem [{\citenamefont {Dutta}\ \emph {et~al.}(2001)\citenamefont {Dutta},
  \citenamefont {Reno}, \citenamefont {Sarcevic},\ and\ \citenamefont
  {Seckel}}]{Dutta:2000hh}%
  \BibitemOpen
  \bibfield  {author} {\bibinfo {author} {\bibfnamefont {S.~I.}\ \bibnamefont
  {Dutta}}, \bibinfo {author} {\bibfnamefont {M.~H.}\ \bibnamefont {Reno}},
  \bibinfo {author} {\bibfnamefont {I.}~\bibnamefont {Sarcevic}},\ and\
  \bibinfo {author} {\bibfnamefont {D.}~\bibnamefont {Seckel}},\ }\href
  {https://doi.org/10.1103/PhysRevD.63.094020} {\bibfield  {journal} {\bibinfo
  {journal} {Phys. Rev. D}\ }\textbf {\bibinfo {volume} {63}},\ \bibinfo
  {pages} {094020} (\bibinfo {year} {2001})},\ \Eprint
  {https://arxiv.org/abs/hep-ph/0012350} {arXiv:hep-ph/0012350} \BibitemShut
  {NoStop}%
\bibitem [{\citenamefont {Ehlert}\ \emph {et~al.}(2024)\citenamefont {Ehlert},
  \citenamefont {van Vliet}, \citenamefont {Oikonomou},\ and\ \citenamefont
  {Winter}}]{Ehlert:2023btz}%
  \BibitemOpen
  \bibfield  {author} {\bibinfo {author} {\bibfnamefont {D.}~\bibnamefont
  {Ehlert}}, \bibinfo {author} {\bibfnamefont {A.}~\bibnamefont {van Vliet}},
  \bibinfo {author} {\bibfnamefont {F.}~\bibnamefont {Oikonomou}},\ and\
  \bibinfo {author} {\bibfnamefont {W.}~\bibnamefont {Winter}},\ }\href
  {https://doi.org/10.1088/1475-7516/2024/02/022} {\bibfield  {journal}
  {\bibinfo  {journal} {JCAP}\ }\textbf {\bibinfo {volume} {02}},\ \bibinfo
  {pages} {022}},\ \Eprint {https://arxiv.org/abs/2304.07321} {arXiv:2304.07321
  [astro-ph.HE]} \BibitemShut {NoStop}%
\bibitem [{\citenamefont {Aab}\ \emph {et~al.}(2016)\citenamefont {Aab} \emph
  {et~al.}}]{PierreAuger:2016qzd}%
  \BibitemOpen
  \bibfield  {author} {\bibinfo {author} {\bibfnamefont {A.}~\bibnamefont
  {Aab}} \emph {et~al.} (\bibinfo {collaboration} {Pierre Auger}),\ }\href@noop
  {} {\  (\bibinfo {year} {2016})},\ \Eprint {https://arxiv.org/abs/1604.03637}
  {arXiv:1604.03637 [astro-ph.IM]} \BibitemShut {NoStop}%
\bibitem [{\citenamefont {Feldman}\ and\ \citenamefont
  {Cousins}(1998)}]{Feldman:1997qc}%
  \BibitemOpen
  \bibfield  {author} {\bibinfo {author} {\bibfnamefont {G.~J.}\ \bibnamefont
  {Feldman}}\ and\ \bibinfo {author} {\bibfnamefont {R.~D.}\ \bibnamefont
  {Cousins}},\ }\href {https://doi.org/10.1103/PhysRevD.57.3873} {\bibfield
  {journal} {\bibinfo  {journal} {Phys. Rev. D}\ }\textbf {\bibinfo {volume}
  {57}},\ \bibinfo {pages} {3873} (\bibinfo {year} {1998})},\ \Eprint
  {https://arxiv.org/abs/physics/9711021} {arXiv:physics/9711021} \BibitemShut
  {NoStop}%
\bibitem [{\citenamefont {Huang}\ \emph {et~al.}(2023)\citenamefont {Huang},
  \citenamefont {Jana}, \citenamefont {Lindner},\ and\ \citenamefont
  {Rodejohann}}]{Huang:2022pce}%
  \BibitemOpen
  \bibfield  {author} {\bibinfo {author} {\bibfnamefont {G.-Y.}\ \bibnamefont
  {Huang}}, \bibinfo {author} {\bibfnamefont {S.}~\bibnamefont {Jana}},
  \bibinfo {author} {\bibfnamefont {M.}~\bibnamefont {Lindner}},\ and\ \bibinfo
  {author} {\bibfnamefont {W.}~\bibnamefont {Rodejohann}},\ }\href
  {https://doi.org/10.1016/j.physletb.2023.137842} {\bibfield  {journal}
  {\bibinfo  {journal} {Phys. Lett. B}\ }\textbf {\bibinfo {volume} {840}},\
  \bibinfo {pages} {137842} (\bibinfo {year} {2023})},\ \Eprint
  {https://arxiv.org/abs/2204.10347} {arXiv:2204.10347 [hep-ph]} \BibitemShut
  {NoStop}%
\bibitem [{\citenamefont {Magill}\ \emph {et~al.}(2018)\citenamefont {Magill},
  \citenamefont {Plestid}, \citenamefont {Pospelov},\ and\ \citenamefont
  {Tsai}}]{Magill:2018jla}%
  \BibitemOpen
  \bibfield  {author} {\bibinfo {author} {\bibfnamefont {G.}~\bibnamefont
  {Magill}}, \bibinfo {author} {\bibfnamefont {R.}~\bibnamefont {Plestid}},
  \bibinfo {author} {\bibfnamefont {M.}~\bibnamefont {Pospelov}},\ and\
  \bibinfo {author} {\bibfnamefont {Y.-D.}\ \bibnamefont {Tsai}},\ }\href
  {https://doi.org/10.1103/PhysRevD.98.115015} {\bibfield  {journal} {\bibinfo
  {journal} {Phys. Rev. D}\ }\textbf {\bibinfo {volume} {98}},\ \bibinfo
  {pages} {115015} (\bibinfo {year} {2018})},\ \Eprint
  {https://arxiv.org/abs/1803.03262} {arXiv:1803.03262 [hep-ph]} \BibitemShut
  {NoStop}%
\bibitem [{\citenamefont {Mu{\~n}oz-Ovalle}\ and\ \citenamefont
  {Vogl}(2026)}]{Munoz-Ovalle:2025riq}%
  \BibitemOpen
  \bibfield  {author} {\bibinfo {author} {\bibfnamefont {A.}~\bibnamefont
  {Mu{\~n}oz-Ovalle}}\ and\ \bibinfo {author} {\bibfnamefont {S.}~\bibnamefont
  {Vogl}},\ }\href {https://doi.org/10.1103/6f5j-5j7l} {\bibfield  {journal}
  {\bibinfo  {journal} {Phys. Rev. D}\ }\textbf {\bibinfo {volume} {113}},\
  \bibinfo {pages} {055004} (\bibinfo {year} {2026})},\ \Eprint
  {https://arxiv.org/abs/2510.02450} {arXiv:2510.02450 [hep-ph]} \BibitemShut
  {NoStop}%
\bibitem [{\citenamefont {Barducci}\ and\ \citenamefont
  {Dondarini}(2024)}]{Barducci:2024kig}%
  \BibitemOpen
  \bibfield  {author} {\bibinfo {author} {\bibfnamefont {D.}~\bibnamefont
  {Barducci}}\ and\ \bibinfo {author} {\bibfnamefont {A.}~\bibnamefont
  {Dondarini}},\ }\href {https://doi.org/10.1007/JHEP10(2024)165} {\bibfield
  {journal} {\bibinfo  {journal} {JHEP}\ }\textbf {\bibinfo {volume} {10}},\
  \bibinfo {pages} {165 (2024)}},\ \Eprint {https://arxiv.org/abs/2404.09609}
  {arXiv:2404.09609 [hep-ph]} \BibitemShut {NoStop}%
\bibitem [{\citenamefont {Miranda}\ \emph {et~al.}(2021)\citenamefont
  {Miranda}, \citenamefont {Papoulias}, \citenamefont {Sanders}, \citenamefont
  {T{\'o}rtola},\ and\ \citenamefont {Valle}}]{Miranda:2021kre}%
  \BibitemOpen
  \bibfield  {author} {\bibinfo {author} {\bibfnamefont {O.~G.}\ \bibnamefont
  {Miranda}}, \bibinfo {author} {\bibfnamefont {D.~K.}\ \bibnamefont
  {Papoulias}}, \bibinfo {author} {\bibfnamefont {O.}~\bibnamefont {Sanders}},
  \bibinfo {author} {\bibfnamefont {M.}~\bibnamefont {T{\'o}rtola}},\ and\
  \bibinfo {author} {\bibfnamefont {J.~W.~F.}\ \bibnamefont {Valle}},\ }\href
  {https://doi.org/10.1007/JHEP12(2021)191} {\bibfield  {journal} {\bibinfo
  {journal} {JHEP}\ }\textbf {\bibinfo {volume} {12}},\ \bibinfo {pages} {191
  (2021)}},\ \Eprint {https://arxiv.org/abs/2109.09545} {arXiv:2109.09545
  [hep-ph]} \BibitemShut {NoStop}%
\end{thebibliography}%
